\documentstyle[aps,epsfig]{revtex}

\newcommand{\beq}{\begin{equation}}
\newcommand{\eeq}{\end{equation}}
\newcommand{\ba}{\begin{eqnarray}}
\newcommand{\ea}{\nonumber \end{eqnarray}}
\newcommand{\be}{\begin{eqnarray}}
\newcommand{\ee}{\nonumber \end{eqnarray}}

\def\fun#1#2{\lower3.6pt\vbox{\baselineskip0pt\lineskip.9pt
\ialign{$\mathsurround=0pt#1\hfil##\hfil$\crcr#2\crcr\sim\crcr}}}

\def\wh{\widehat}

\begin{document}

\title{Systematics of quark--antiquark states:
where are the lightest  glueballs?}
\author{V.V. Anisovich}
\date{Talk given at HADRON-2003, 31 August -- 5 September 2003,
Aschaffensburg, Germany}
\maketitle

\begin{abstract}

The analysis of the experimental data of Crystal Barrel Collaboration on
the $p\bar p$ annihilation in flight with the production of mesons in the
final state resulted in a discovery of a large number of mesons over
the region  1900--2400~MeV, thus allowing us to systematize
quark-antiquark states in the
$(n,M^2)$  and $(J,M^2)$ planes, where $n$ and $J$ are radial quantum
number and spin of the meson with the mass $M$. The data point to meson
trajectories in these planes being approximately linear, with a
universal slope.
Basing on these data and results of the recent K-matrix analysis
a nonet classification is performed. In the
scalar-isoscalar sector, the broad resonance
state $f_0(1200-1600)$ is superfluous for
the $q\bar q$ classification, i.e. it is an
exotic state. The ratios of coupling constants for the transitions
$f_0\to \pi\pi, K\bar K, \eta\eta,\eta\eta'$
point to the gluonium nature of the broad state $f_0(1200-1600)$.
The problem of the location of the lightest pseudoscalar glueball is
also discussed.
\end{abstract}
\maketitle

 $\,$\\
The search for exotic mesons should be based on the classification of
$q\bar q$-states. Exotic mesons are those which are superfluous for
the $q\bar q$ systematics. The quark--antiquark systematics means:\\
(i) classification of $q\bar q$ states as states located on the
$(n,M^2)$ and $(J,M^2)$ trajectories, and\\
(ii) determination of the quark--gluonium content of states from
the analysis of the decay coupling constants, namely, hadronic
and radiative decay couplings as well as weak ones.

For the hadronic decay coupling constants, the most reliable information
comes from the $K$-matrix analysis. In addition, the $K$-matrix
analysis allows us to study  bare states (the states before the onset
of the decay processes).
$\,$\\

{\bf 1.Systematics of the $q\bar q$-states on the
$(n,M^2)$ and $(J,M^2)$ planes.}   \\
The analysis of experimental data on the $p\bar p$ annihilation in
flight with the production of mesons in the final state resulted in a
discovery of the large number of mesons over the region 1900--2400
MeV \cite{ral}. This allowed us to systematize quark--antiquark
states on the $(n,M^2)$ and $(J,M^2)$ planes. The data point to almost
the linear meson trajectories on these planes, with a universal slope
\cite{syst}.

In Fig. 1, one can see the $(n,M^2)$ trajectories for the $(I=1)$
states, which are
drown for the $a_1$- and $a_3$-mesons (Fig.
1a), $\pi$-, $\pi_2$- and $\pi_4$-mesons (Fig. 1b), $b_1$- and
$b_3$-mesons (Fig. 1c). All these trajectories reveal linear behaviour,
such as
$$
M^2=M_0^2+\mu^2(n-1),
\nonumber
$$
with $\mu^2\simeq 1.2$ GeV$^2$; $M_0$ is the mass of the ground (basic)
state, $n=1$.  The pion,
being beyond the trajectory, is an exception, that is not a
surprise, for the pion is a special particle in certain respect. In the
classification, all these mesons should be treated as $q\bar
q$-states. Using the spectroscopy notations for $q\bar q$-states,
$^{2S+1}L_J$ where
$S$ is the quark spin and $L$ is their orbital momentum,
we assign the trajectories
to mesons as follows: \\
$\,$\\
$ a_1(1230)\,{\rm trajectory}$: $n^3P_1q\bar q$-states,
$ a_3(2030)\,{\rm trajectory}$: $ n^3F_3q\bar q$-states; \\
$\,$ $ \pi(140)\,{\rm trajectory}$:  $n^1S_0q\bar q $-states,
$ \pi_2(1670)\,{\rm trajectory}$:  $ n^1D_2q\bar q$-states,
$ \pi_4(2250)\,{\rm trajectory}$: $ n^1G_4q\bar q $-states;\\
$\,$ $ b_1(1235)\,{\rm trajectory}$: $ n^1P_1q\bar q $-states,
$ b_3(2020)\,{\rm trajectory}$: $ n^1F_3q\bar q $-states.\\
$\,$\\
In Fig. 1c the state $b_1(1640)$ is shown  which was not
discovered in the experiment but is predicted by trajectories:
the states we predict are denoted by open circles.

The trajectories
$\rho,\rho_3,a_0,a_2,a_4$ demonstrate linear behaviour as well. The
 $\rho$-trajectories are shown in Fig. 2a:\\
$\,$\\
$\rho (770)$ trajectory:
 dominantly $n^3S_1\, q\bar q(L=0)$,
$\quad$ $\rho (1700)$ trajectory:
 dominantly $n ^3D_1\;q\bar q(L=4)$.\\
$\,$\\
The states $ ^3S_1\;q\bar q$ and $ ^3D_1\;q\bar q$
may mix with each other but considerable mass splitting of the
$\rho (770)$ and $\rho (1700)$ states tells us that the mixing
is not large.
The $\rho_3$
and $a_0$, $a_2$, $a_4$ trajectories
are shown in Figs. 2b and 2c:\\
$\,$\\
$\rho_3(1690)$ trajectory: dominantly
$n ^3D_3\;q\bar q(L=2)$,  $\quad$
$\rho_3(2260)$ trajectory:  dominantly $n ^3G_3\;q\bar q(L=4)$,\\
$a_0(980)$ trajectory: $n ^3P_0\;q\bar q(L=1)$, $\quad$
$a_2(1320)$ trajectory:
dominantly $n^3P_2\;q\bar q(L=1)$,\\
$a_2(2030)$ trajectory:  dominantly
$n ^3F_2\;q\bar q(L=3)$, $\quad$
$a_4(2005)$ trajectory:   dominantly $n ^3F_4\;q\bar q(L=3)$.\\

\newpage

\begin{figure}[h]
%Fig. 1
\centerline{\epsfig{file=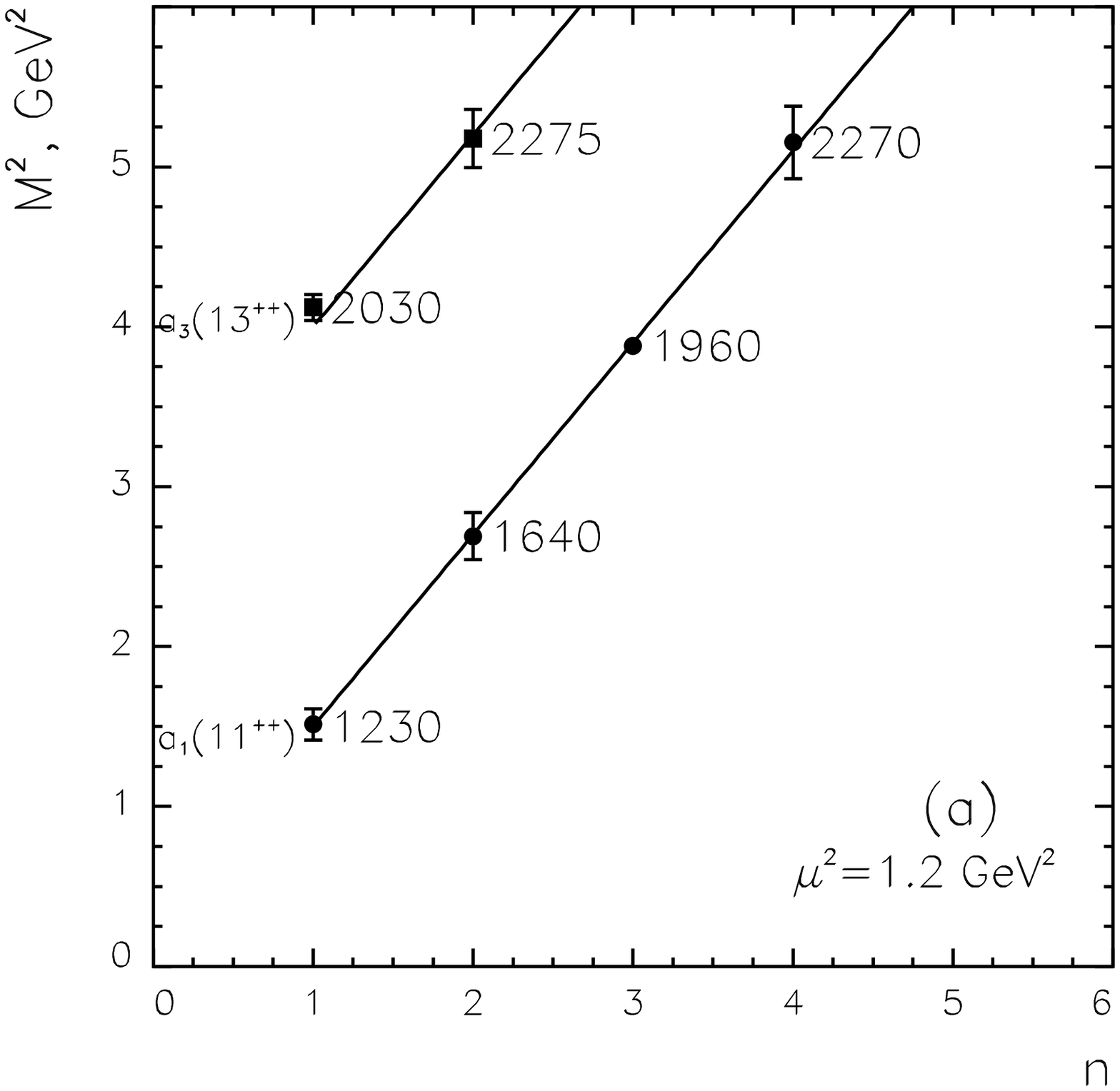,width=7cm}\hspace{-1.5cm}
            \epsfig{file=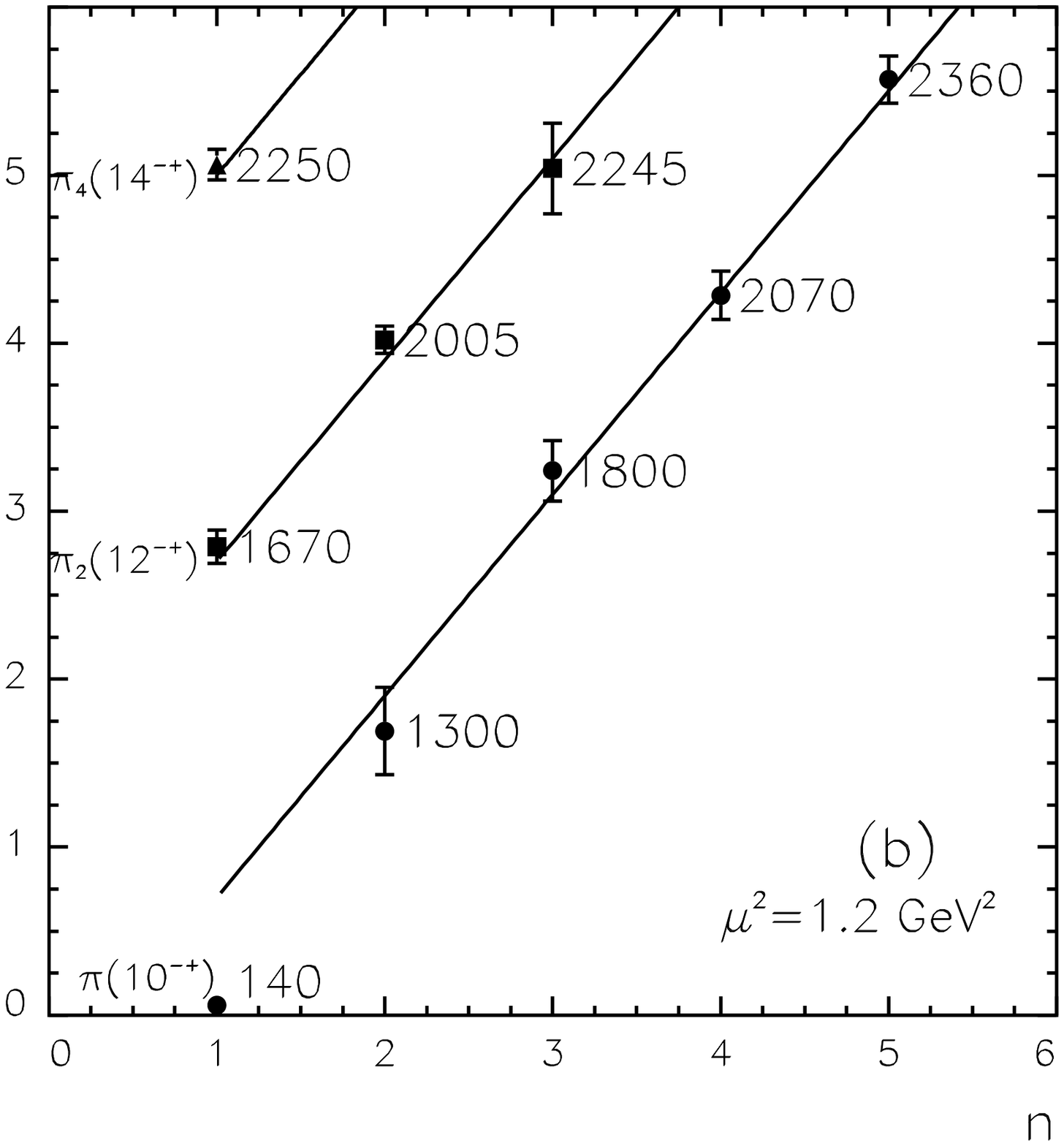,width=7cm}\hspace{-1.5cm}
            \epsfig{file=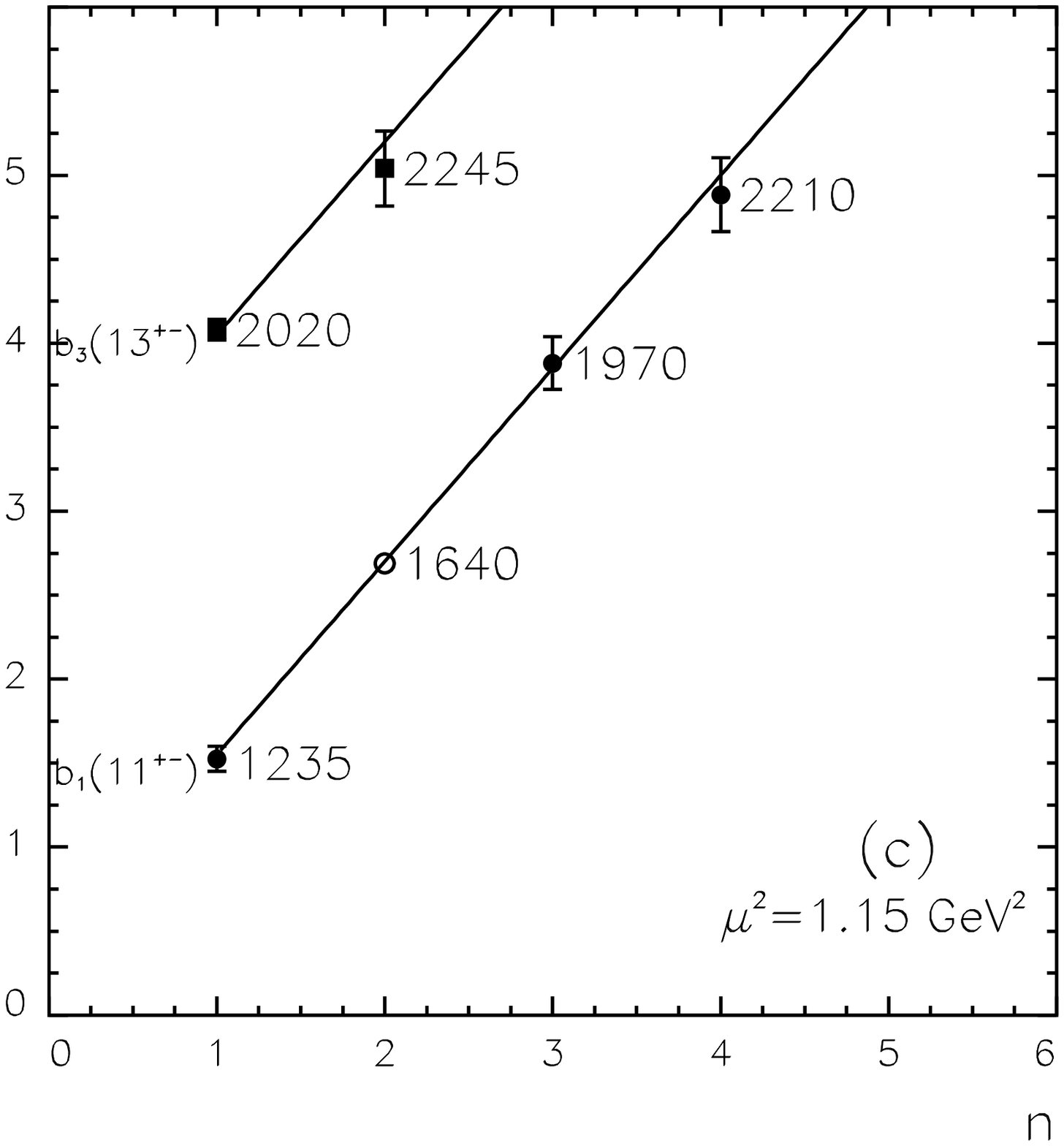,width=7cm}}
\vspace{-0.5cm}
\caption{The (I=1)-mesons on the
$(n,M^2)$ planes:  $\pi, a_1, a_3, b_1,b_3$ rajectories  }
\end{figure}

%Fig. 2
\begin{figure}[h]
\centerline{\epsfig{file=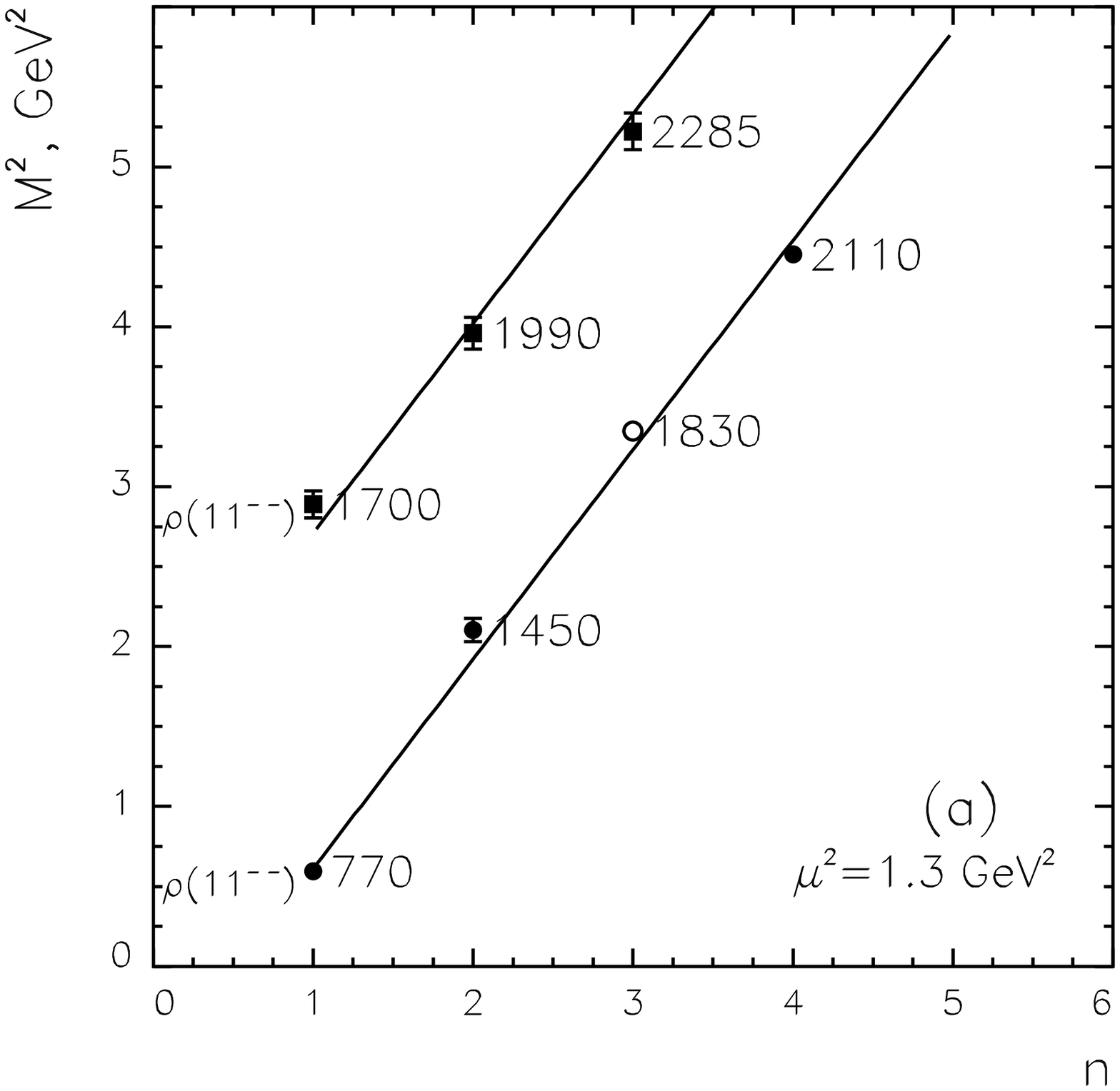,width=7cm}\hspace{-1.5cm}
            \epsfig{file=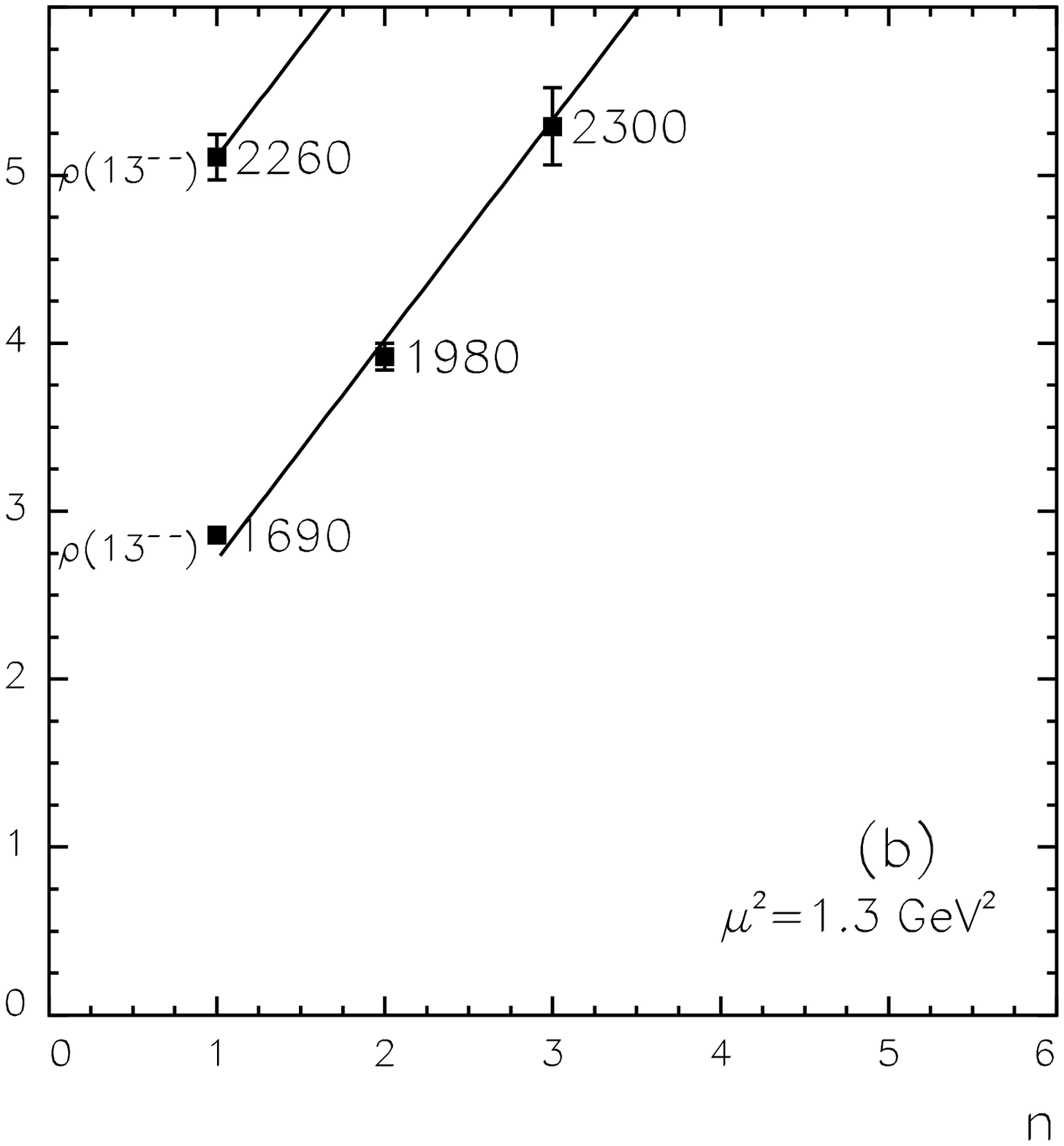,width=7cm}\hspace{-1.5cm}
            \epsfig{file=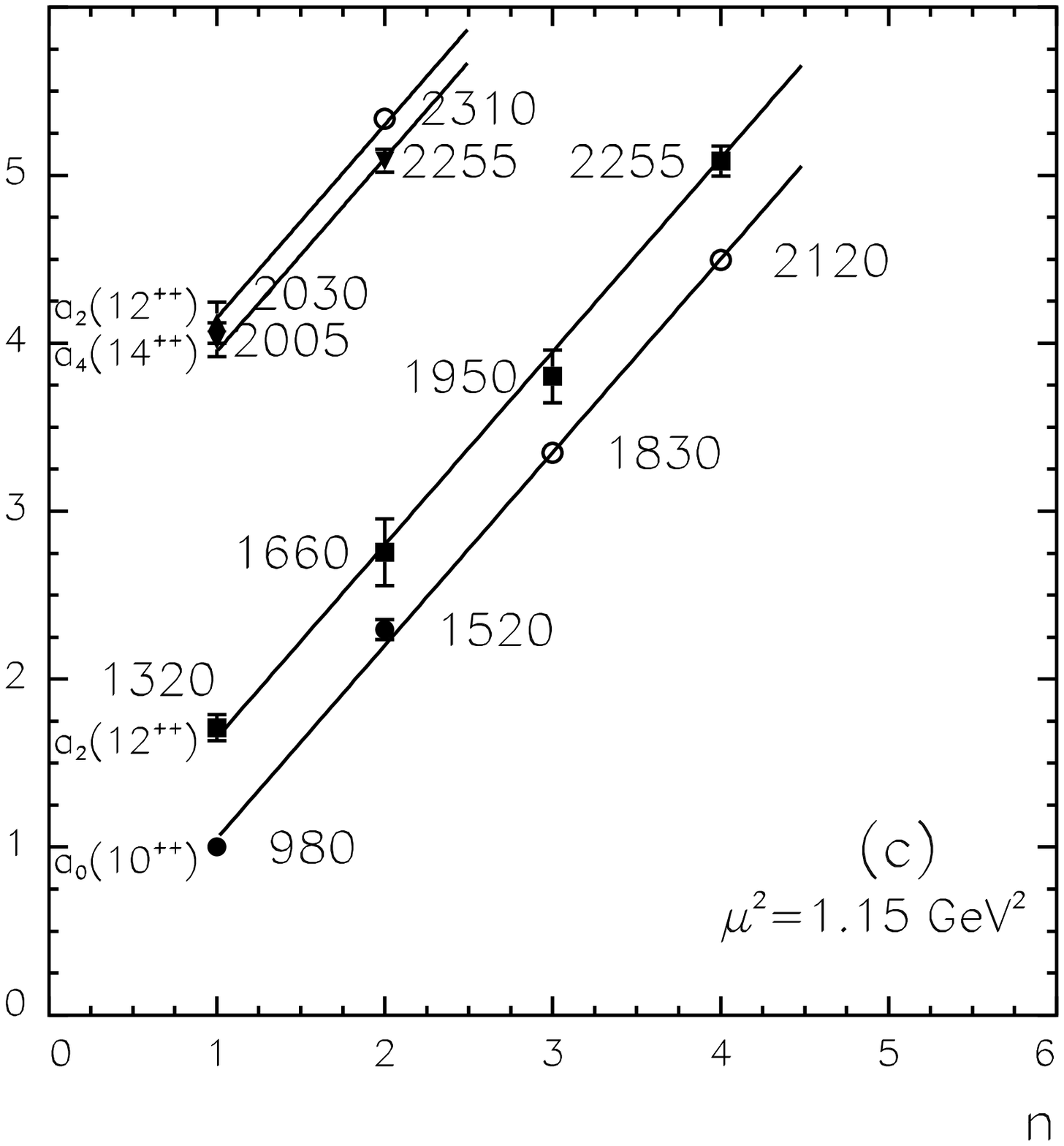,width=7cm}}
\vspace{-0.5cm}
\caption{The (I=1)-mesons on the
$(n,M^2)$ planes: the $\rho,\rho_3, a_0,a_2$ trajectories}
\end{figure}

\begin{figure}[h]
%Fig. 3
\vspace{-0.5cm}
\centerline{\epsfig{file=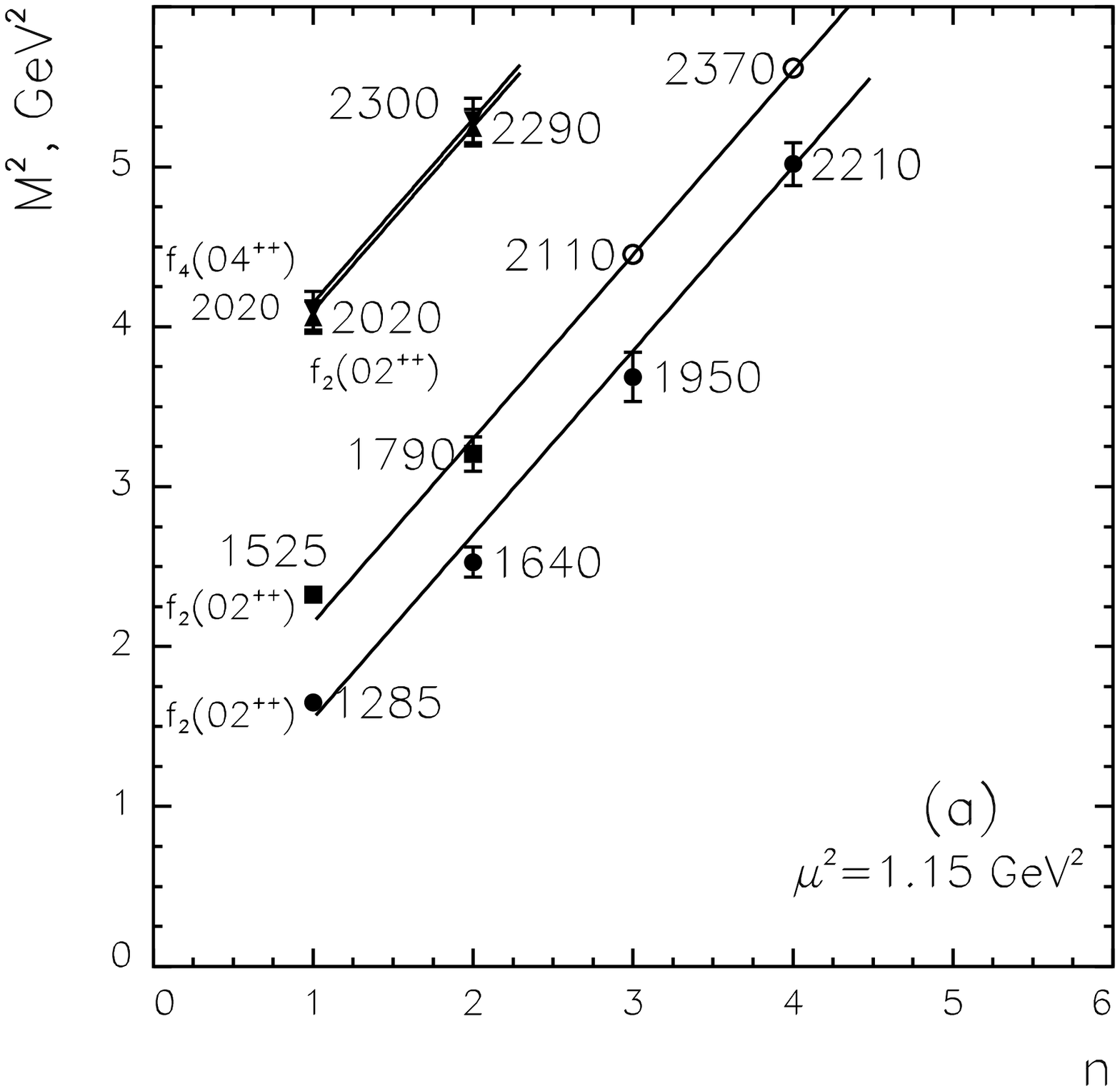,width=7cm}\hspace{-1.5cm}
            \epsfig{file=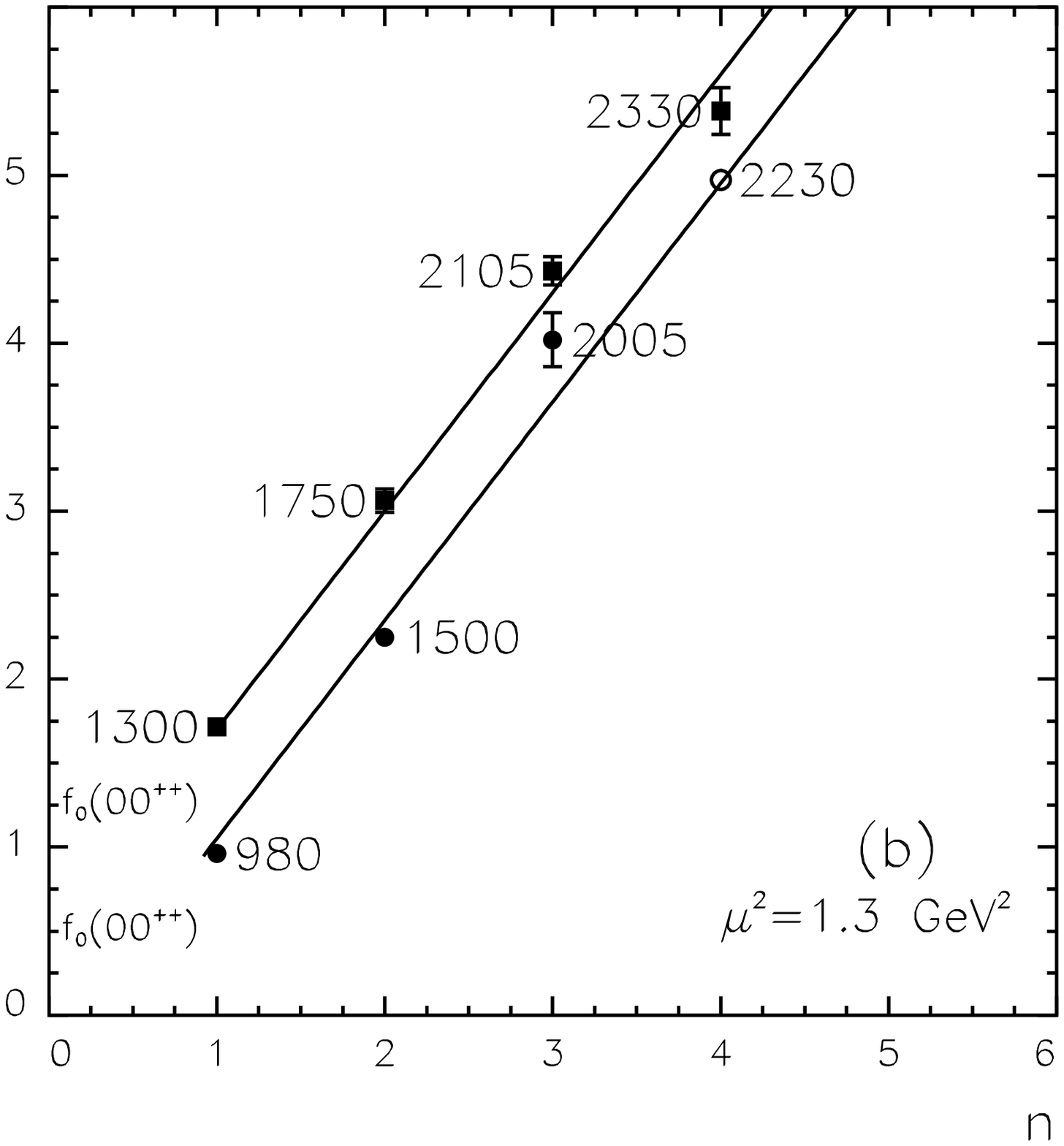,width=7cm}\hspace{-1.5cm}
            \epsfig{file=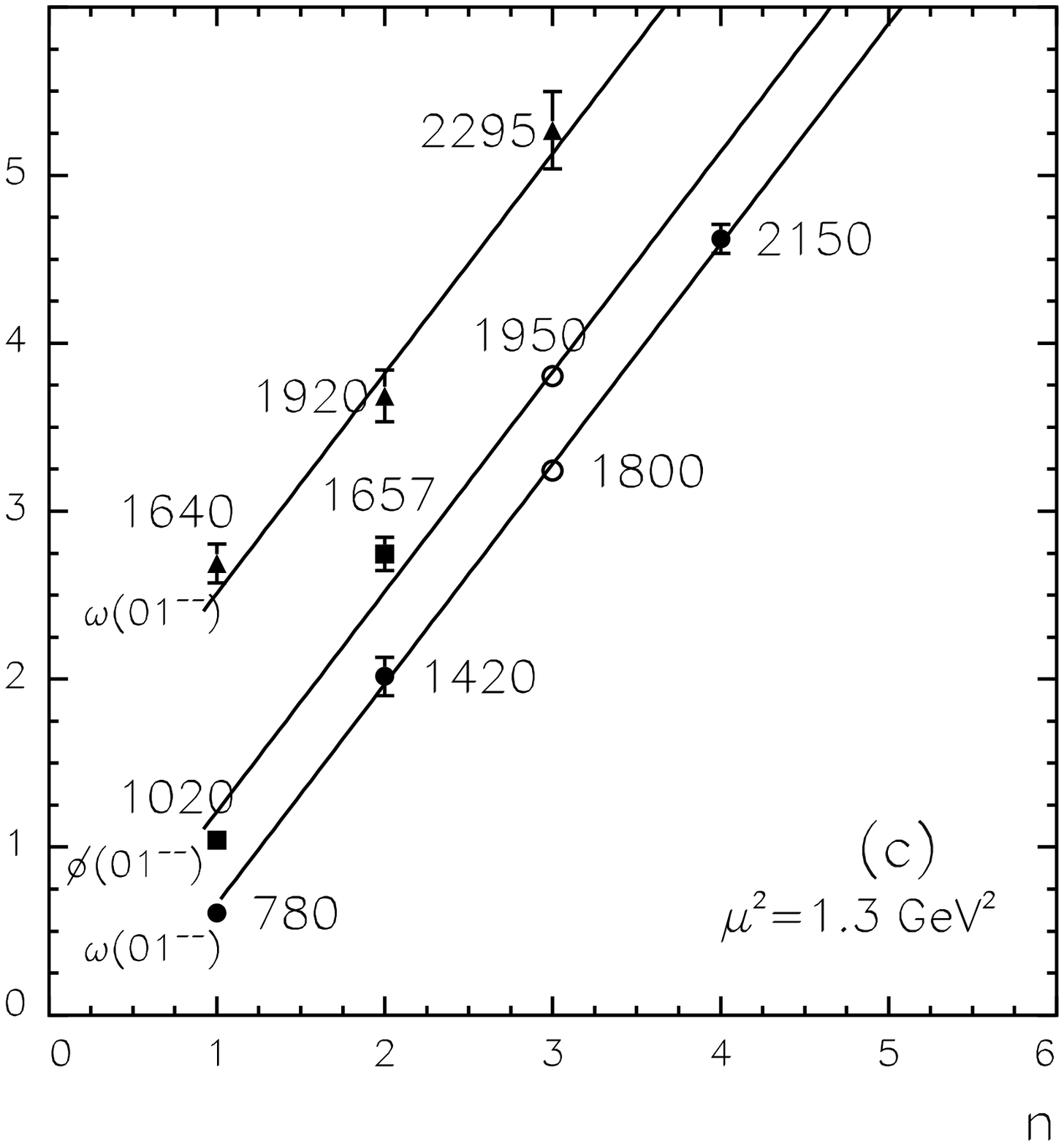,width=7cm}}
\vspace{-0.5cm}
\caption{The (I=0)-states on the $(n,M^2)$ planes:
$f_0,f_2,\omega$ trajectories}

\vspace{-0.6cm}
\end{figure}

\newpage
The $a_0(980)$-resonance, which sometimes is
discussed as a non-$q\bar q$ state, lays on the $q\bar q$-trajectory,
that is an argument in favour of its quark--antiquark origin.

The figures 3 and 4 for the $I=0$ states confirm the
linearity of trajectories on the $(n,M^2)$-planes, with a universal
slope $\mu^2\simeq 1.2-1.3$ GeV$^2$. In this sector, we face
a doubling of trajectories  due to the existence of
 two flavour states $n\bar n=(u\bar u+d\bar d)/\sqrt{2}$ and
$s\bar s$.

In Fig. 3, one can see the  $f_2(1285)$ and $f_2(1525)$ trajectories,
which are dominantly the $ ^3P_2\;(n\bar n +s\bar s)$ states, and
 $f_2(2020)$
and $f_4(2020)$ ones,
which are dominantly $ ^3F_2\;(n\bar n +s\bar s) $
and $ ^3F_4\;(n\bar n +s\bar s) $, correspondingly.
Figure 3b demonstrates
$f_0(980)$ and $f_0(1300)$ trajectories ($ ^3P_0\;(n\bar n +s\bar s) $
states),
while   figure 3c shows the trajectories $\omega (780)$, $\phi (1020)$
(dominantly $ ^3S_1\;(n\bar n +s\bar s) $) and  $\omega (1640)$
(dominantly $ ^3D_1\;(n\bar n +s\bar s) $).

\begin{figure}[h]
%Fig. 4
\centerline{\epsfig{file=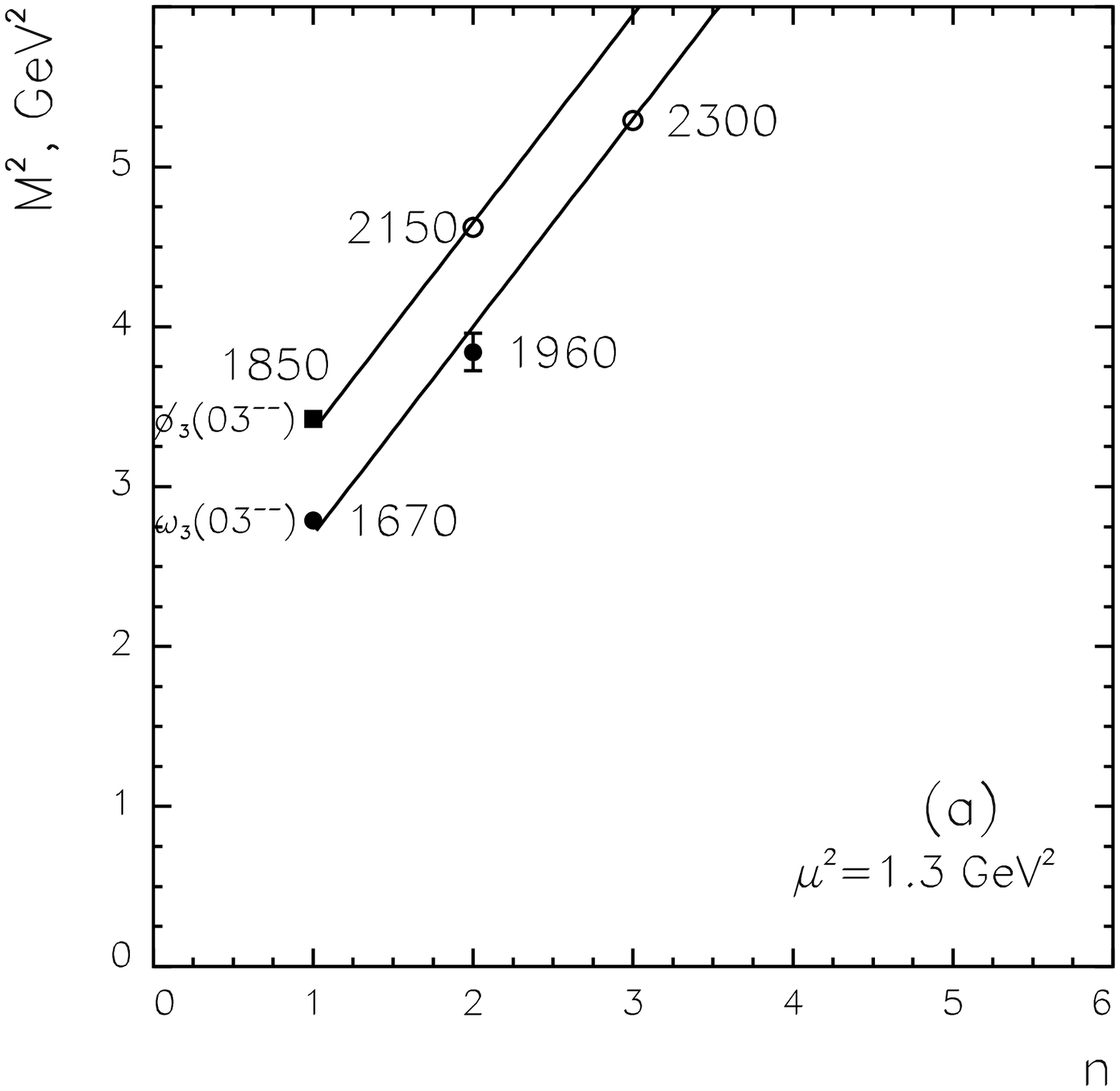,width=7cm}\hspace{-1.5cm}
            \epsfig{file=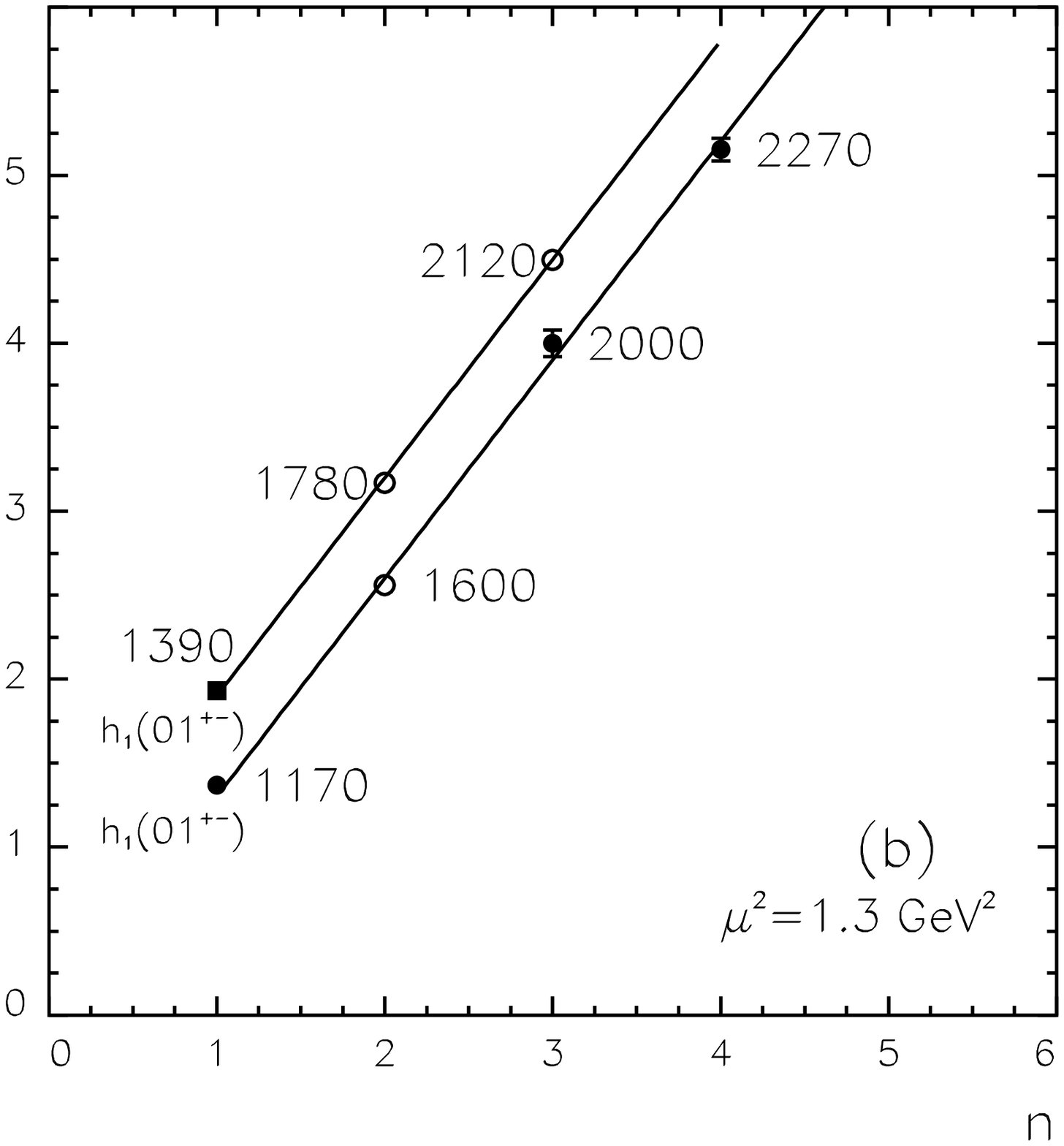,width=7cm}\hspace{-1.5cm}
            \epsfig{file=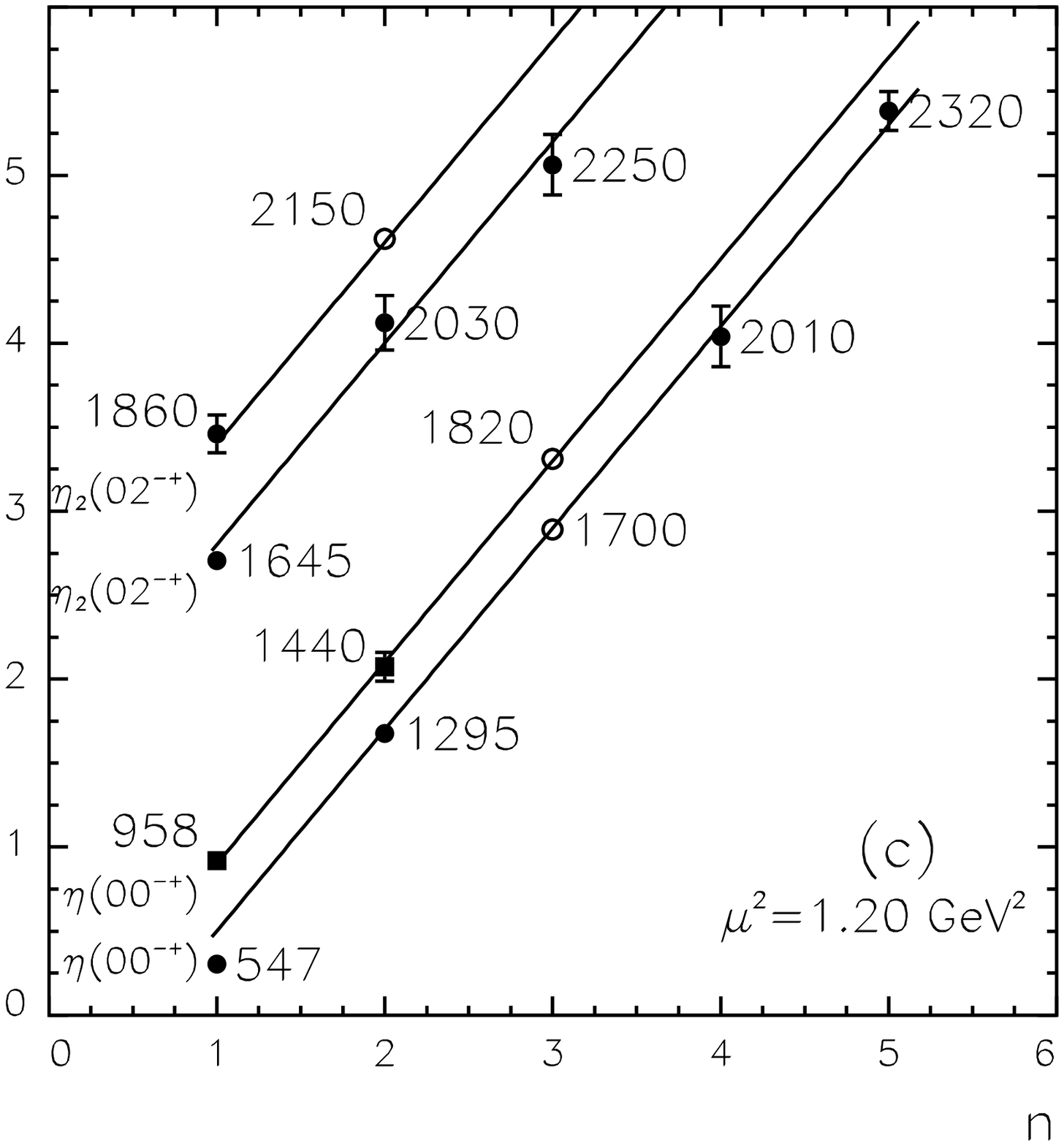,width=7cm}}
\vspace{-0.5cm}
\caption{ The (I=0)-states on the $(n,M^2)$ planes:
$\omega_3,h_1,\eta$ trajectories }
\vspace{-0.1cm}
\end{figure}

\begin{figure}[h]
%Fig. 5
\centerline{\epsfig{file=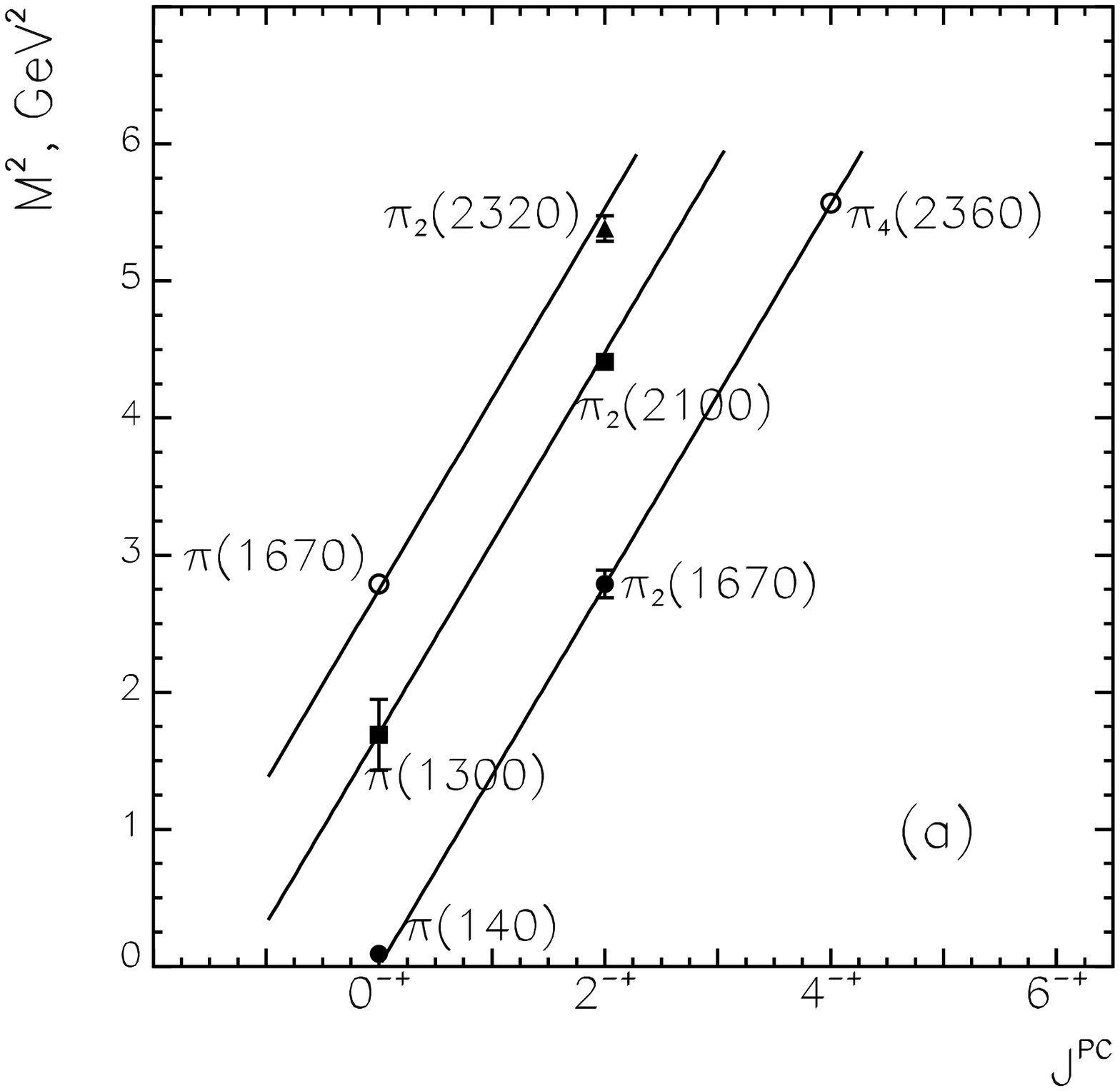,width=7cm}\hspace{-1.5cm}
            \epsfig{file=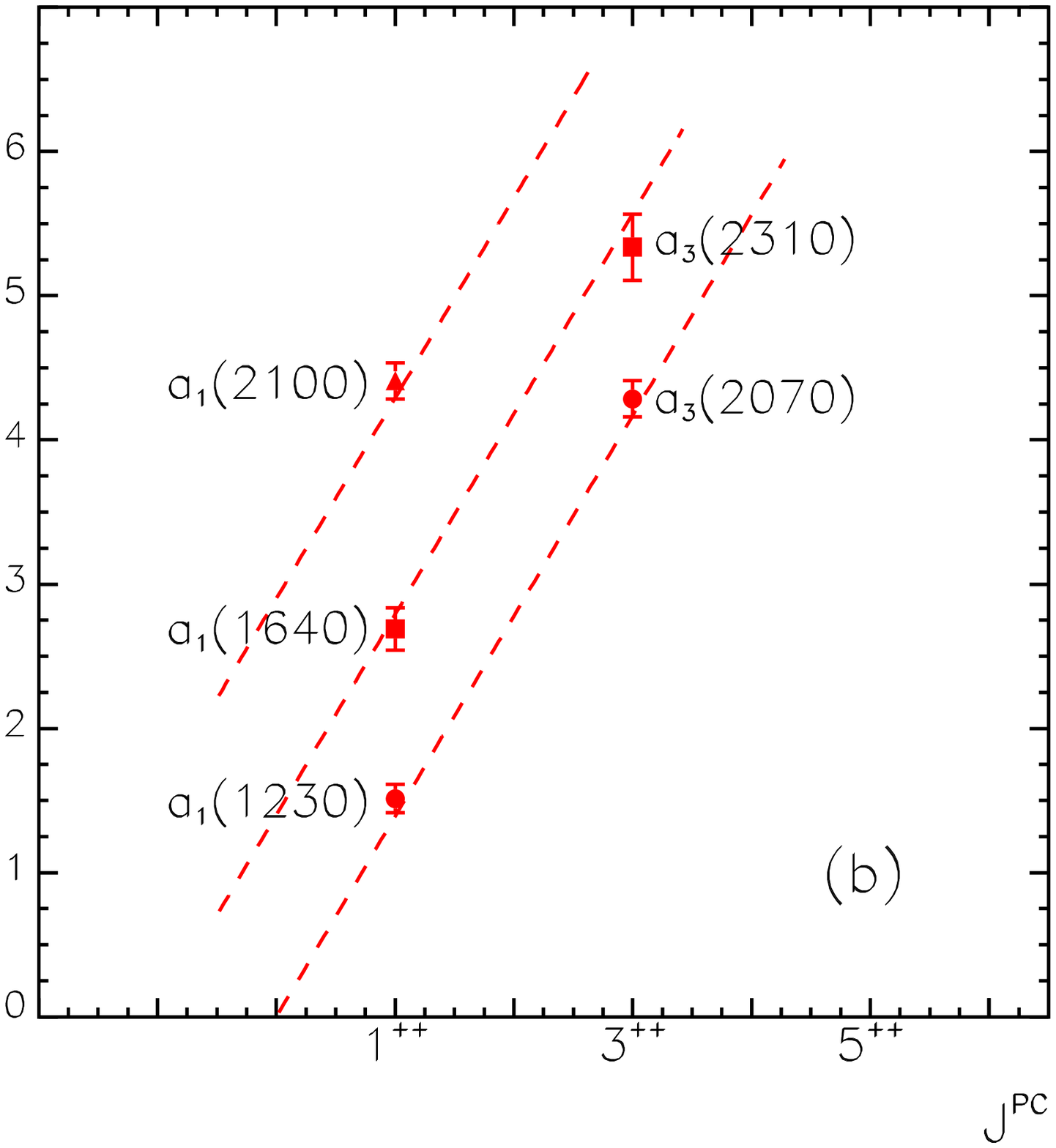,width=7cm}\hspace{-1.5cm}
            \epsfig{file=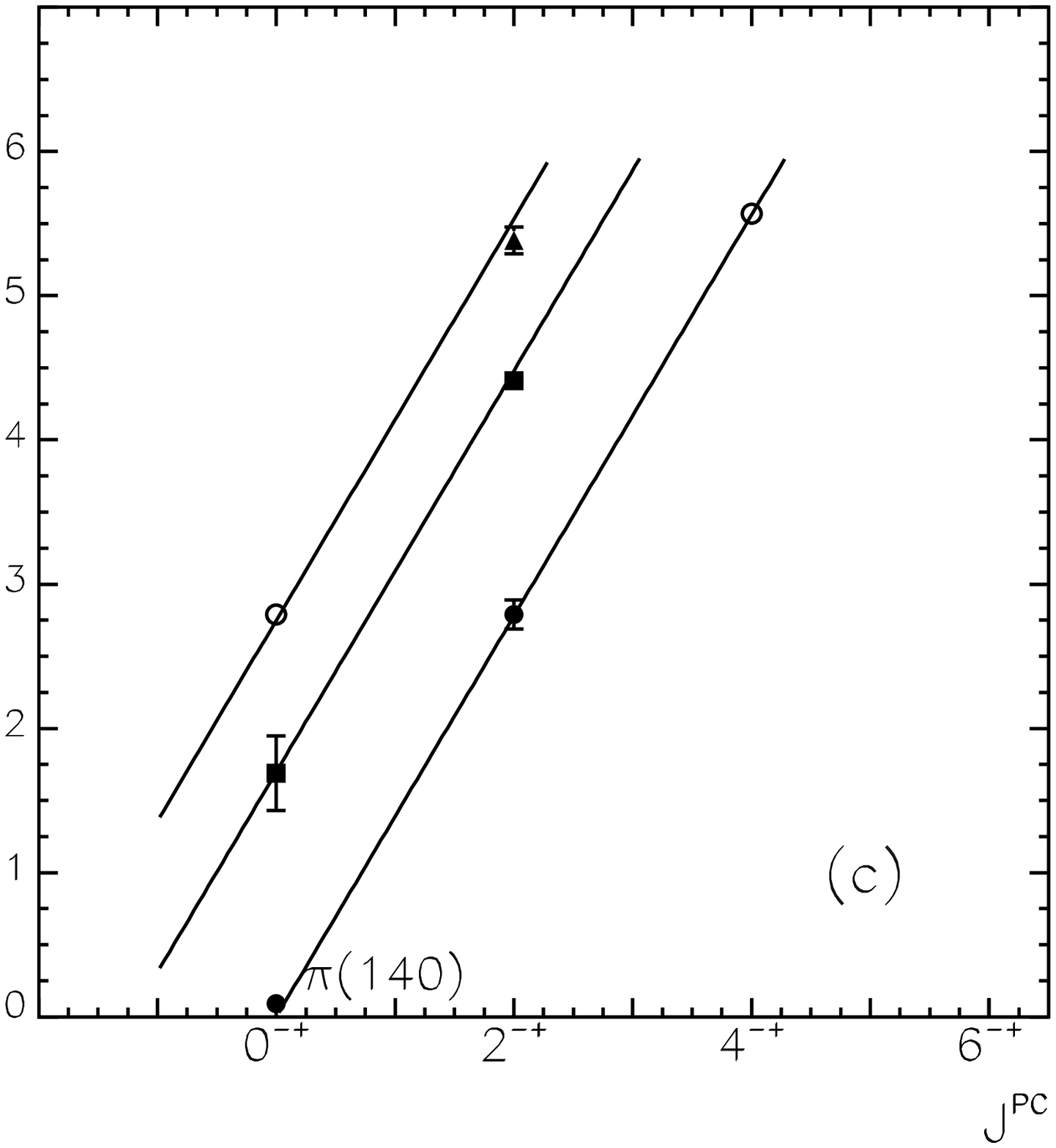,width=7cm}\hspace{-7.13cm}
            \epsfig{file=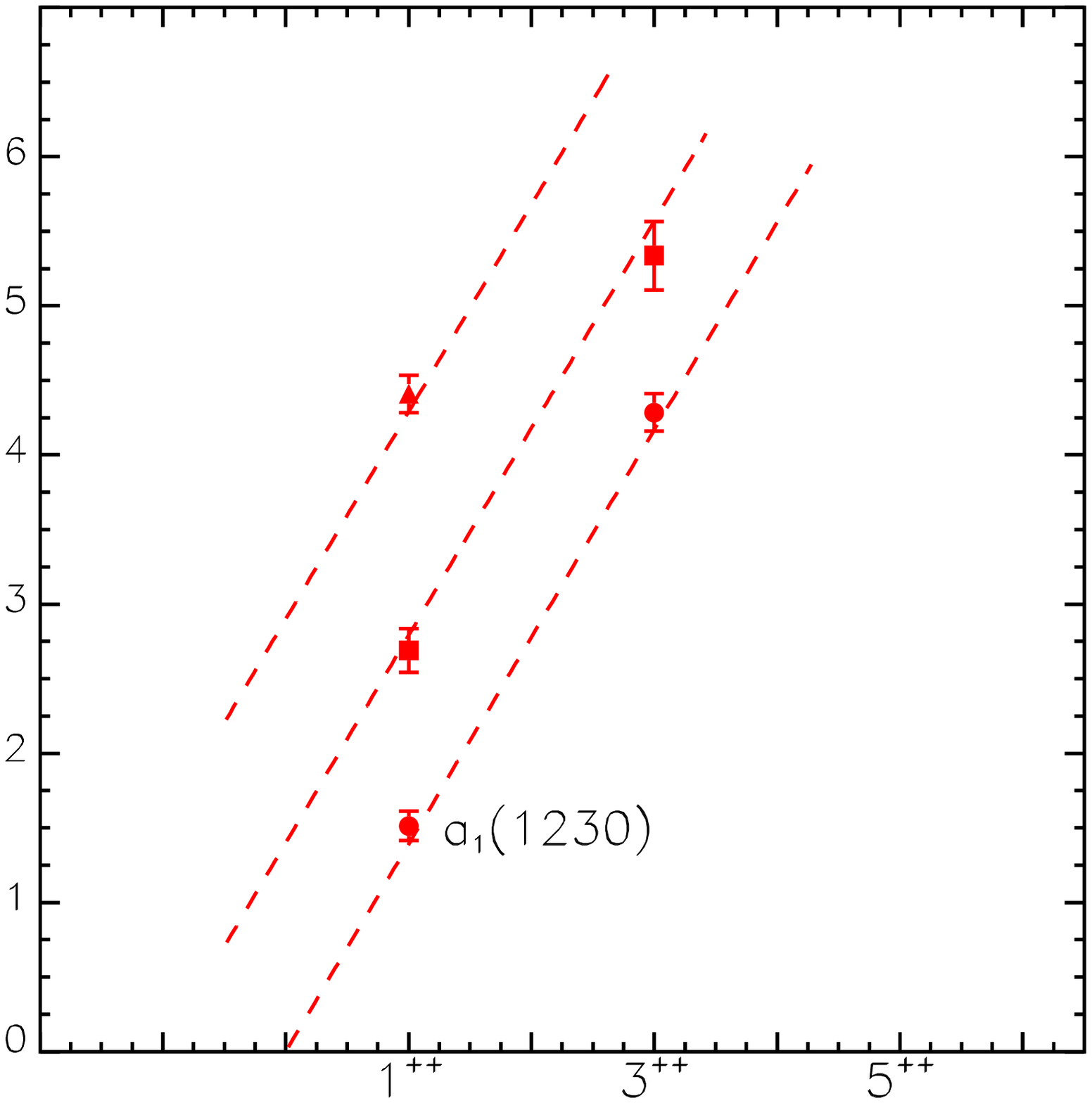,width=7cm}}
\caption{The $\pi$ and $a_1$ trajectories on the $(J,M^2)$ plane.}
\end{figure}

Let me stress that
$f_0(980)$, $f_0(1300)$ (the latter denoted as $f_0(1370)$ in the
compilation \cite{PDG}), $f_0(1500)$, $f_0(1750)$, which are sometimes
discussed as candidates to exotics, lay quite comfortably on the
linear $q\bar q$ trajectories. The  K-matrix analysis \cite{EPJA} gives
 us one more state in the
$(IJ^{PC}=00^{++})$-sector, that is, a broad resonance
$f_0(1200-1600)$: just this state may be considered as exotics, a
descendant of the lightest scalar glueball. It is discussed in the
next section. The light $\sigma$, if it exists, is also beyond the
$q\bar q$ trajectories being in this way a candidate for exotics as
well.

In Figs. 4a,b,c,  one can see trajectoties as follows:
$\omega_3,\; \phi_3$ (dominantely $ ^3D_3\;q\bar q $),
$h_1$ ($ ^1P_1\;q\bar q $), $\eta$ ($ ^1S_0\;q\bar q $)
and $\eta_2$ ($ ^1D_2\;q\bar q $).
Let me emphasize that the situation in the $0^{-+}$ sector is more
complicated than it is seen from Fig. 4c. On the one hand,  the states
$\eta(1440)$ and $\eta(1295)$ lay good enough on the linear $(n,M^2)$
trajectory. On the other hand, experimental
indications for $\eta(1295)$ are not convincing,
and the resonance $\eta(1440)$ reveals itself in different reactions

\begin{figure}[h]
%Fig. 6
\centerline{\epsfig{file=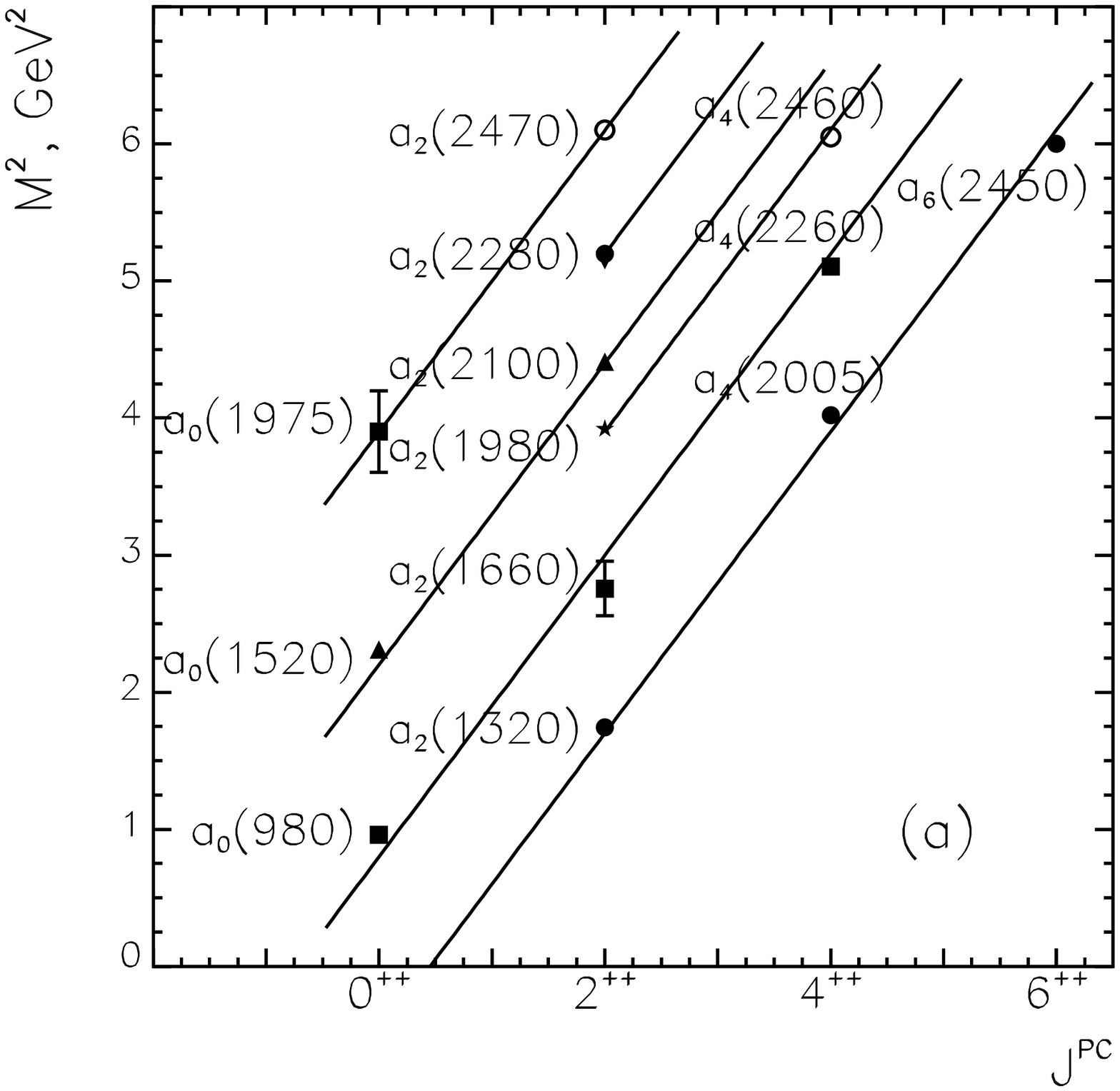,width=7cm}\hspace{-1.5cm}
            \epsfig{file=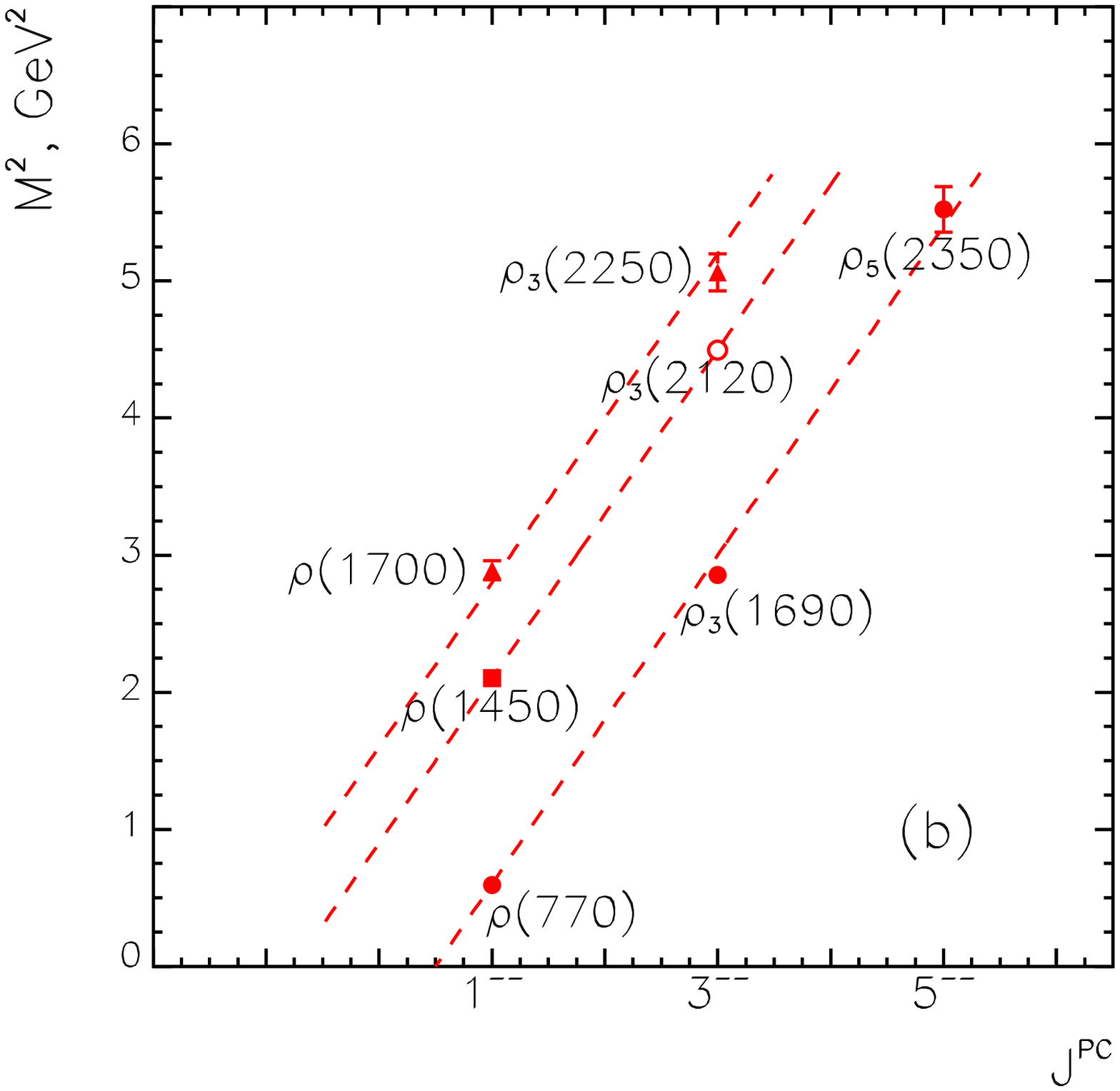,width=7cm}\hspace{-1.5cm}
            \epsfig{file=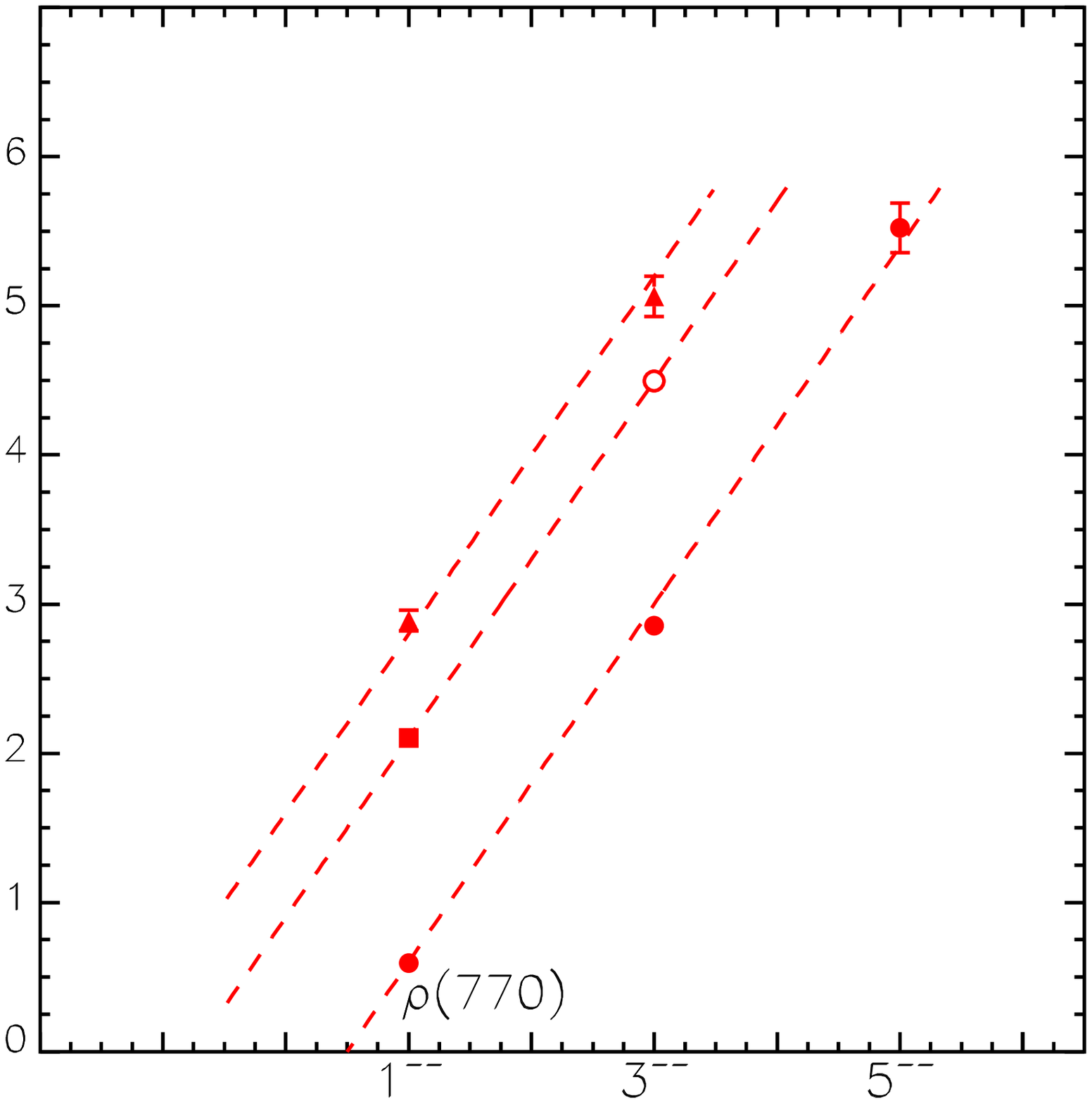,width=7cm}\hspace{-7.13cm}
            \epsfig{file=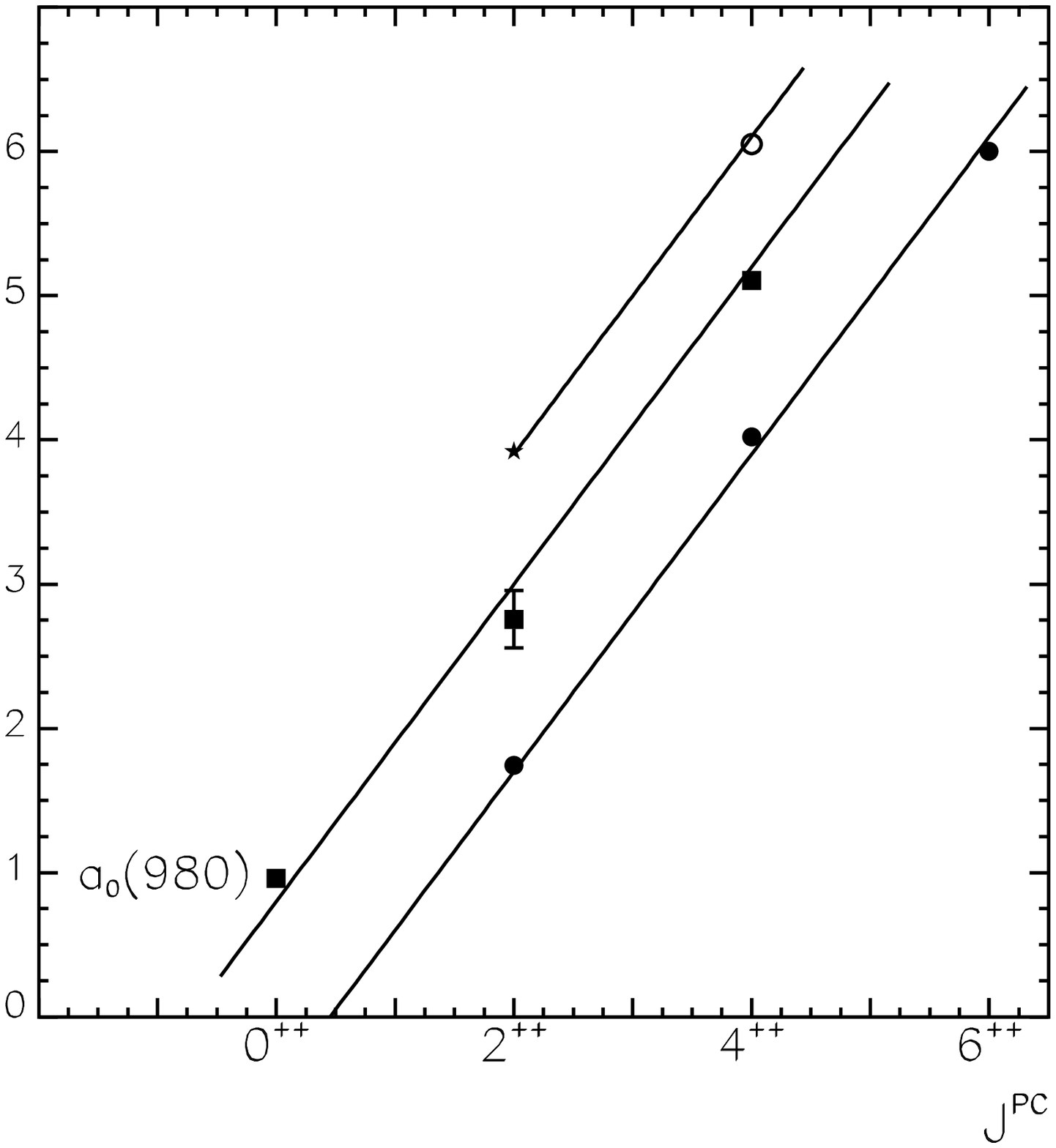,width=7cm}}
\caption{The $\rho$ and $a_2$ trajectories on the $(J,M^2)$ plane.}
\end{figure}

\begin{figure}[h]
%Fig. 7
\centerline{\epsfig{file=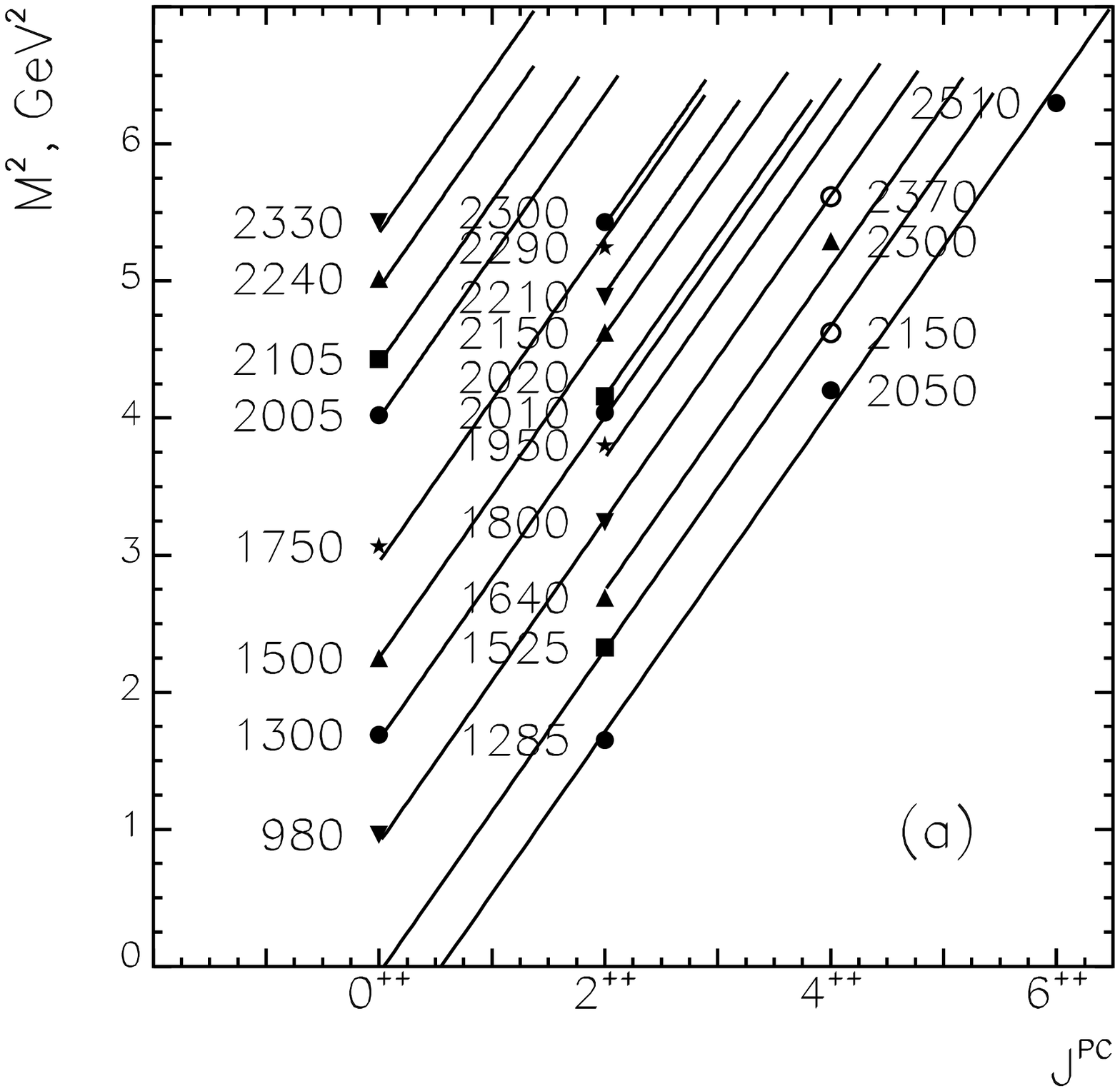,width=7cm}\hspace{-1.5cm}
            \epsfig{file=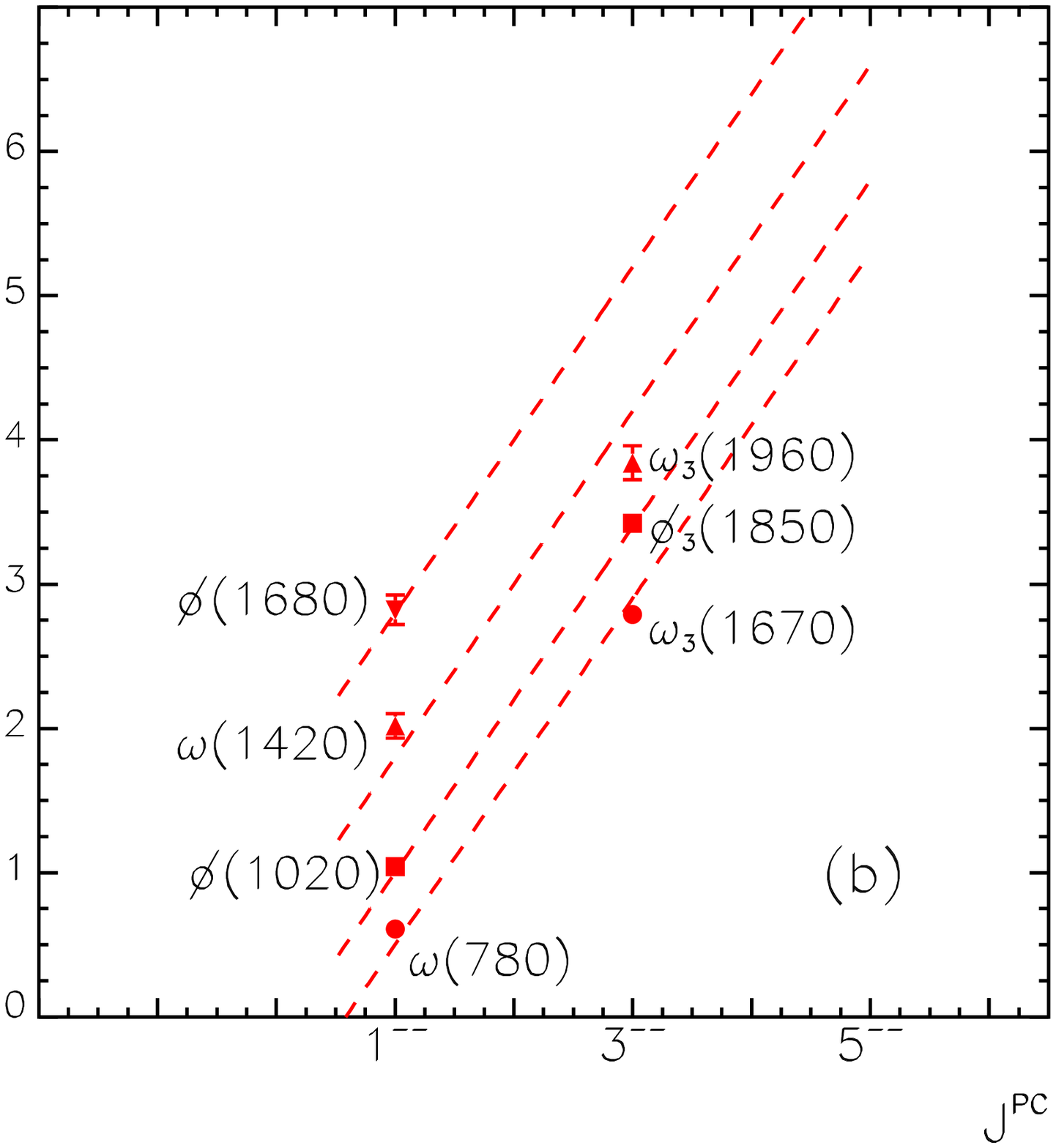,width=7cm}\hspace{-1.5cm}
            \epsfig{file=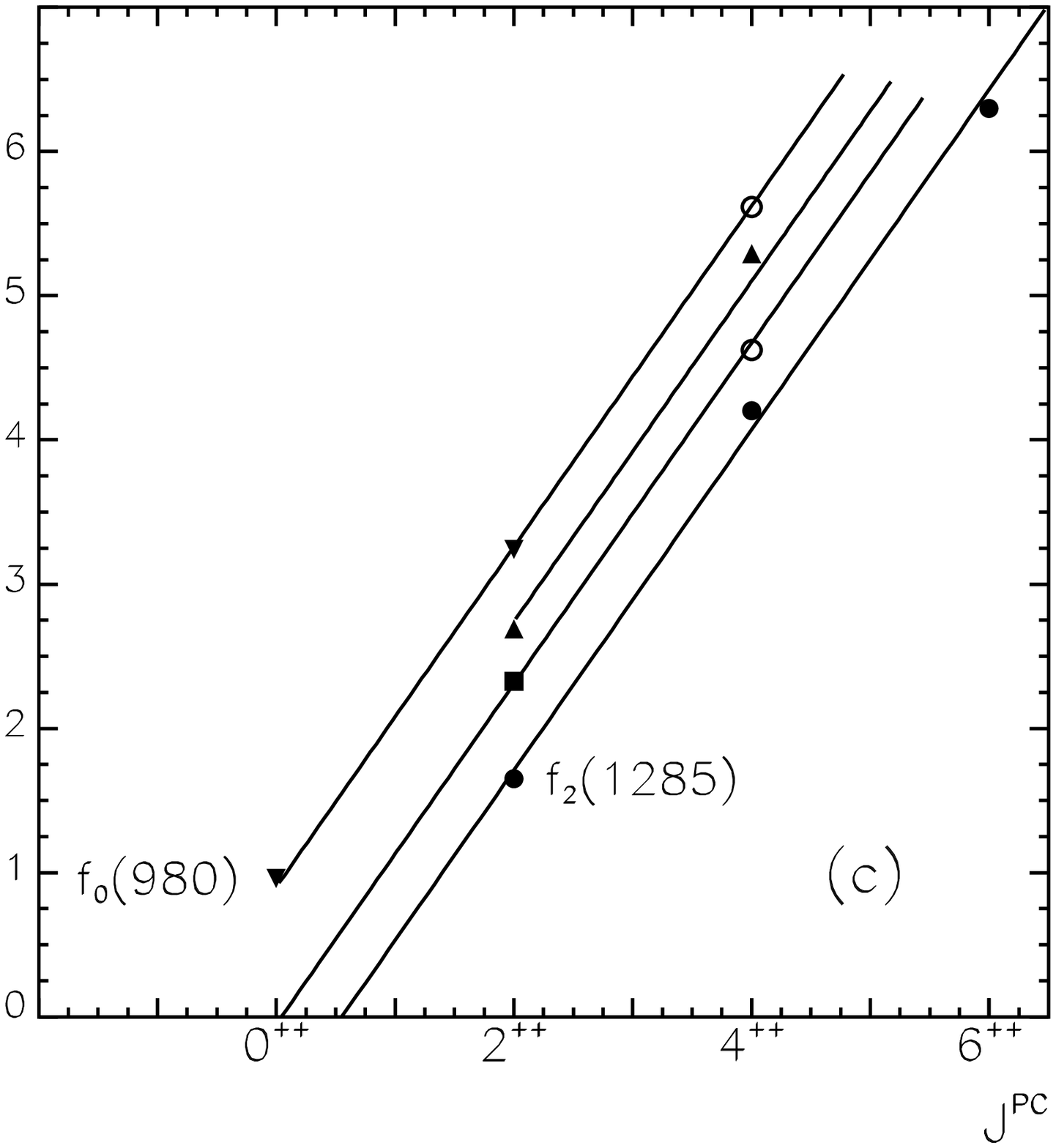,width=7cm}\hspace{-7.13cm}
            \epsfig{file=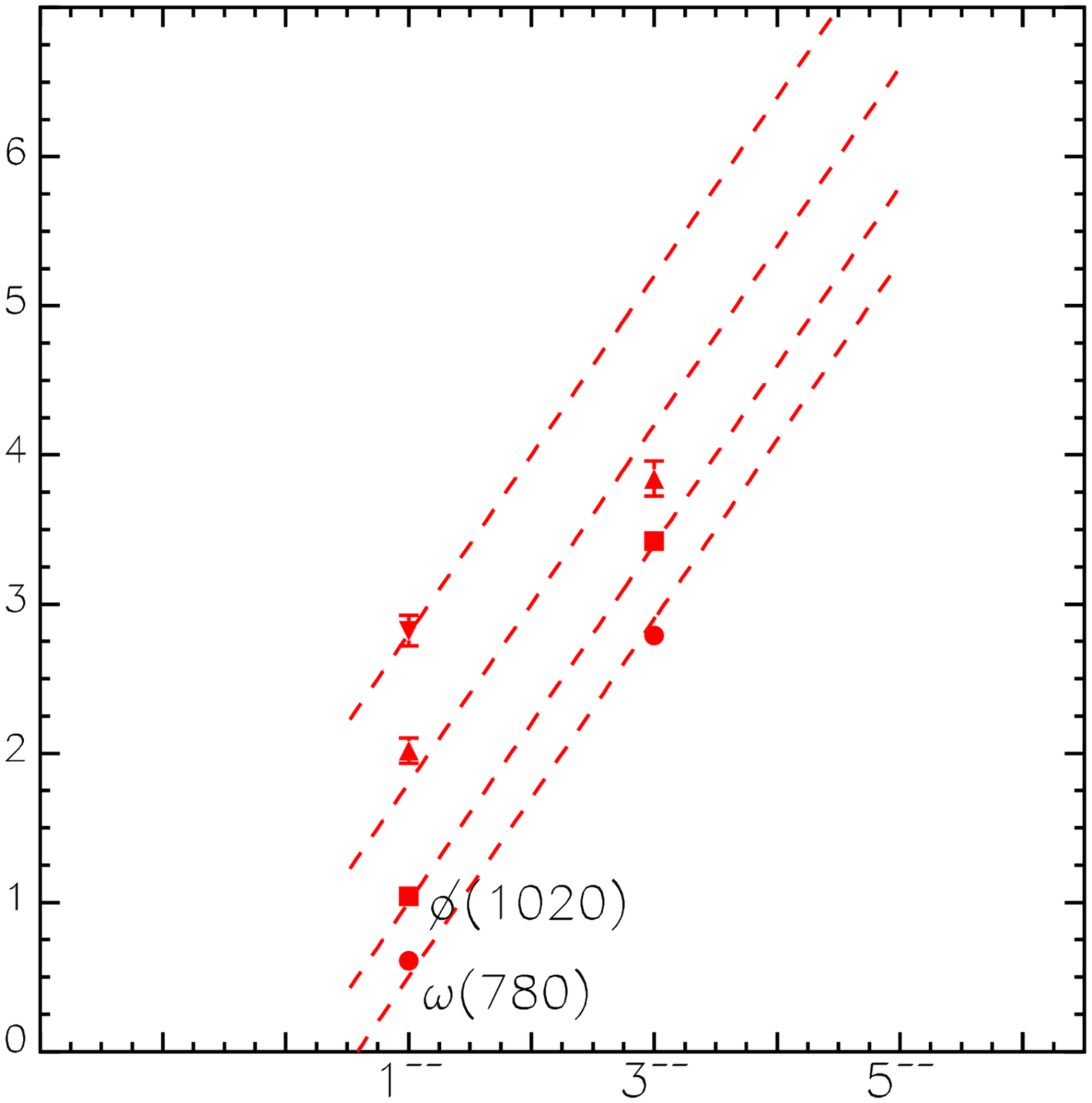,width=7cm}}
\vspace{-0.5cm}
\caption{The $f_2$ and $\omega$ trajectories on the $(J,M^2)$ plane.}
\end{figure}

\noindent
with different
masses: at 1390--1430 MeV in the $\eta\pi\pi$
and $a_0(980)\pi\to K\bar
K \pi$ modes, while in the $K^*(890)\bar K$ mode at
1460--1500 MeV. Besides, in the reaction $J/\psi \to
\gamma\eta(1440)$, the resonance $\eta(1440)$ is produced with a large
background, that may tell us about the existence of a broad state at
1400--1700 MeV. Linear trajectories in the $0^{-+}$
sector predict two
$q\bar q$ states in the neighbouring region, namely, $\eta(1700)$ and
$\eta(1820)$, but these resonance were not seen till now,
because, with available  data
this mass region is difficult for the study.

Now let us discuss the location of mesons on the $(J,M^2)$-planes.
The pion and $a_1$ trajectories are shown in Fig. 5a,b
together with their daughter ones. The trajectories for different
parities are degenerate --- this fact is illustrated by Fig. 5c, where
the combined presentation of $\pi$ and $a_1$ trajectories is given.

The $(J,M^2)$ trajectories for $\rho$ and $a_2$ depicted in Fig. 6
provide us unambigous information on the $a_0(980)$ resonance.  One can
clearly see that $a_0(980)$ lays on the linear daughter trajectory ---
this is  obvious from their combined presentation in Fig.
6c. Supposing that $a_0(980)$ is non-$q\bar q$ meson, one
should expect another $a_0$ state, dominantly $q\bar q$, around 1 GeV.
However, additional $a_0$ state is definitely excluded by the
experimental data.

The $(J,M^2)$ trajectories for the $f_2$ and $\omega$ resonances as
well as for their daughter ones are shown in Fig. 7, and the combined
presentation of Fig. 7c demonstrates the $f_0(980)$ resonance laying on
the daughter $q\bar q$ trajectory.

At last, figure 8 shows us the trajectories for the $K$-meson sector,
the kaons with positive and negative parities lay on the degenerate
trajectories. The $\kappa$ meson discussed as a plausible $0^+$ state
with the mass $\sim 900$ MeV does not belong to linear trajectory, so
it should be considered as an exotic state.

\newpage

\begin{center}
{\bf Table 1. Nonet assignment of the $q\bar q$ states, $n=1$ and $2$.}

\begin{tabular}{|l|cccc|cccc|}
\hline
{\em $q\bar q$}&
\multicolumn{4}{c|}{\em $n=1$} &
\multicolumn{4}{c|}{\em $n=2$} \\ \cline{2-9}
states&I=1&I=0&I=0&I=$1/2$&I=1&I=0&I=0&I=$1/2$\\
\hline
$^1S_0(0^{-+})$&$\pi(140)$   &$\eta(550)$   &$\eta'(958)$&$K(500)$
               &$\pi(1300)$  &$\eta(1295)$  &$\eta(1440)$&$K(1460)$\\
$^3S_1(1^{--})$&$\rho(770)$  &$\omega(780)$ &$\phi(1020)$&$K^*(890)$
               &$\rho(1450)$ &$\omega(1420)$&$\phi(1657)$&$K_1(1680)$\\
\hline
$^3P_0(0^{++})$&$a_0(980)$  &$f_0(980)$ &$f_0(1300)$ &$K_0(1430)$
               &$a_0(1520)$ &$f_0(1500)$&$f_0(1750)$ &$K_0(1850)$\\
$^3P_1(1^{++})$&$a_1(1230)$ &$f_1(1285)$&$f_1(1510)$ &$K_1(1400)$
               &$a_1(1640)$ &           &            &$K_1(1780)$?\\
$^1P_1(1^{+-})$&$b_1(1235)$ &$h_1(1170)$&$h_1(1390)$ &$K_1(1270)$
               &$b_1(1640)$?&$h_1(1600)$?&$h_1(1780)$?&$K_1(1650)$\\
$^3P_2(2^{++})$&$a_2(1320)$ &$f_2(1285)$&$f_2(1525)$ &$K_2(1430)$
               &$a_2(1660)$ &$f_2(1640)$&$f_2(1790)$ &$K_2(1980)$\\
\hline
$^3D_1(1^{--})$&$\rho(1700)$  &$\omega(1640)$  &              &$K_1(1680)$
               &$\rho(1990)$  &$\omega(1920)$  &              &\\
$^3D_2(2^{--})$&              &                &              &$K_2(1800)$
               &              &                &              &$K_2(2170)$?\\
$^1D_2(2^{-+})$&$\pi_2(1670)$ &$\eta_2(1645)$  &$\eta_2(1860)$&$K_2(1580)$
               &$\pi_2(2005)$ &$\eta_2(2030)$  &             &$K_2(2050)$?\\
$^3D_3(3^{--})$&$\rho_3(1690)$&$\omega_3(1670)$&$\phi_3(1850)$&$K_3(1780)$
               &$\rho_3(1980)$&$\omega_3(1960)$&$\phi_3(2150)$?&$K_3(1780)$\\
\hline
$^3F_2(2^{++})$&$a_2(2030)$ &$f_2(2020)$& &
               &$a_2(2310)$?&$f_2(2290)$& &\\
$^3F_3(3^{++})$&$a_3(2030)$ &           & &$K_3(2030)$?
               &$a_3(2275)$ &           & &$K_3(2320)$\\
$^1F_3(3^{+-})$&$b_3(2020)$ &           & &$K_3(1960)$?
               &$b_3(2245)$ &           & &$K_3(2220)$?\\
$^3F_4(4^{++})$&$a_4(2005)$ &$f_4(2020)$& &$K_4(2045)$
               &$a_4(2255)$ &$f_4(2300)$& & \\
\hline
\end{tabular}
\end{center}

\begin{center}
{\bf Table 2. Nonet assignment of $q\bar q$ states, $n=3$ and $4$.}

\begin{tabular}{|l|cccc|cccc|}
\hline
{\em $q\bar q$}&
\multicolumn{4}{c|}{\em $n=3$} &
\multicolumn{4}{c|}{\em $n=4$} \\ \cline{2-9}
states&I=1&I=0&I=0&I=$1/2$&I=1&I=0&I=0&I=1/2\\
\hline
$^1S_0(0^{-+})$&$\pi(1800)$  &$\eta(1700)$?  &$\eta(1820)$?&$K(1830)$
               &$\pi(2070)$  &$\eta(2010)$   &   &\\
$^3S_1(1^{--})$&$\rho(1830)$?&$\omega(1800)$?&$\phi(1950)$?&
               &$\rho(2150)$ &$\omega(2150)$ &    &\\
\hline
$^3P_0(0^{++})$&$a_0(1830)$?&$f_0(2005)$&$f_0(2105)$ &
               &$a_0(2120)$?&$f_0(2230)$?&$f_0(2330)$&   \\
$^3P_1(1^{++})$&$a_1(1960)$ &           &            &
               &$a_1(2270)$ &           &  &\\
$^1P_1(1^{+-})$&$b_1(1970)$ &$h_1(2000)$&$h_1(2120)$?&
               &$b_1(2210)$ &$h_1(2270)$&  & \\
$^3P_2(2^{++})$&$a_2(1950)$ &$f_2(1950)$&$f_2(2110)$?&$K_2(2370)$?
               &$a_2(2255)$ &$f_2(2210)$&$f_2(2370)$?&\\
\hline
$^3D_1(1^{--})$&$\rho(2285)$  &$\omega(2295)$&  &
               &              &                 & &\\
$^3D_2(2^{--})$& &            &                 &$K_2(2400)$?
               &              &                 & &\\
$^1D_2(2^{-+})$&$\pi_2(2245)$ &$\eta_2(2250)$   &&$K_2(2250)$
               &              &                 & &\\
$^3D_3(3^{--})$&$\rho_3(2300)$&$\omega_3(2300)$?&&
               &              &                 &&\\
\hline
\end{tabular}
\end{center}

Summing up the results of meson systematization on the $(n,M^2)$ and
$(J,M^2)$ trajectories, one can state that the resonances
$f_0(980),f_0(1300)$ ($f_0(1370)$ in \cite{PDG}),$f_0(1500)$ and
$f_0(1750)$ ($f_0(1710)$ in \cite{PDG}) are located on
 quark--antiquark trajectories. They originate from the standard
quark--antiquark states.

Now we can assign mesons to the flavour multiplets (see Tables 1,
2 for the multiplets with $n=1,2,3,4$). The interrogation sign
marks the states, which were not discovered by the experiment but are
predicted by linear $(n,M^2)$ and $(J,M^2)$ trajectories. The empty
places are left for the states, which were not discovered and cannot be
predicted.

\begin{figure}[h]
%Fig. 8
\centerline{\epsfig{file=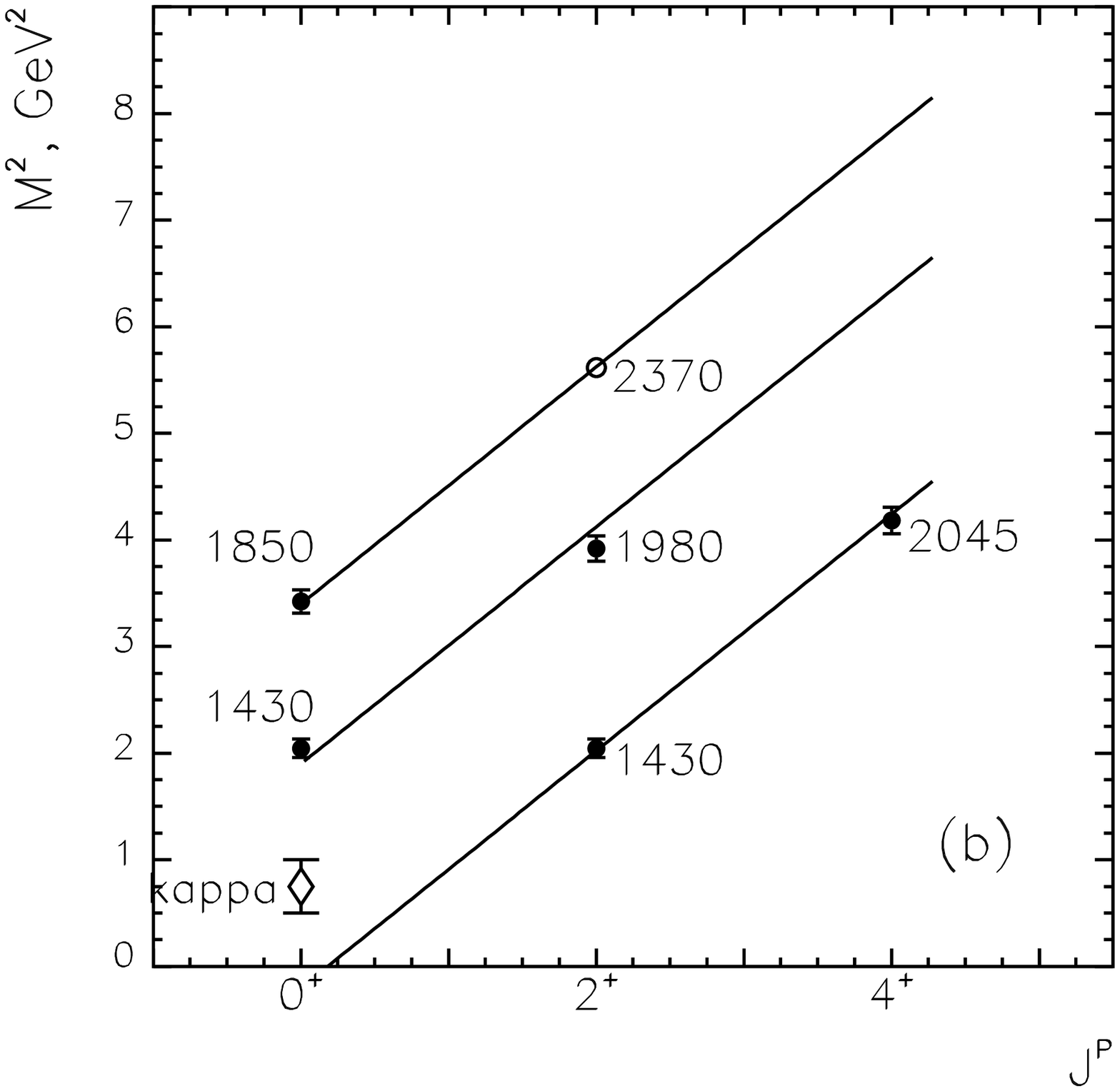,width=7cm}\hspace{-1.5cm}
            \epsfig{file=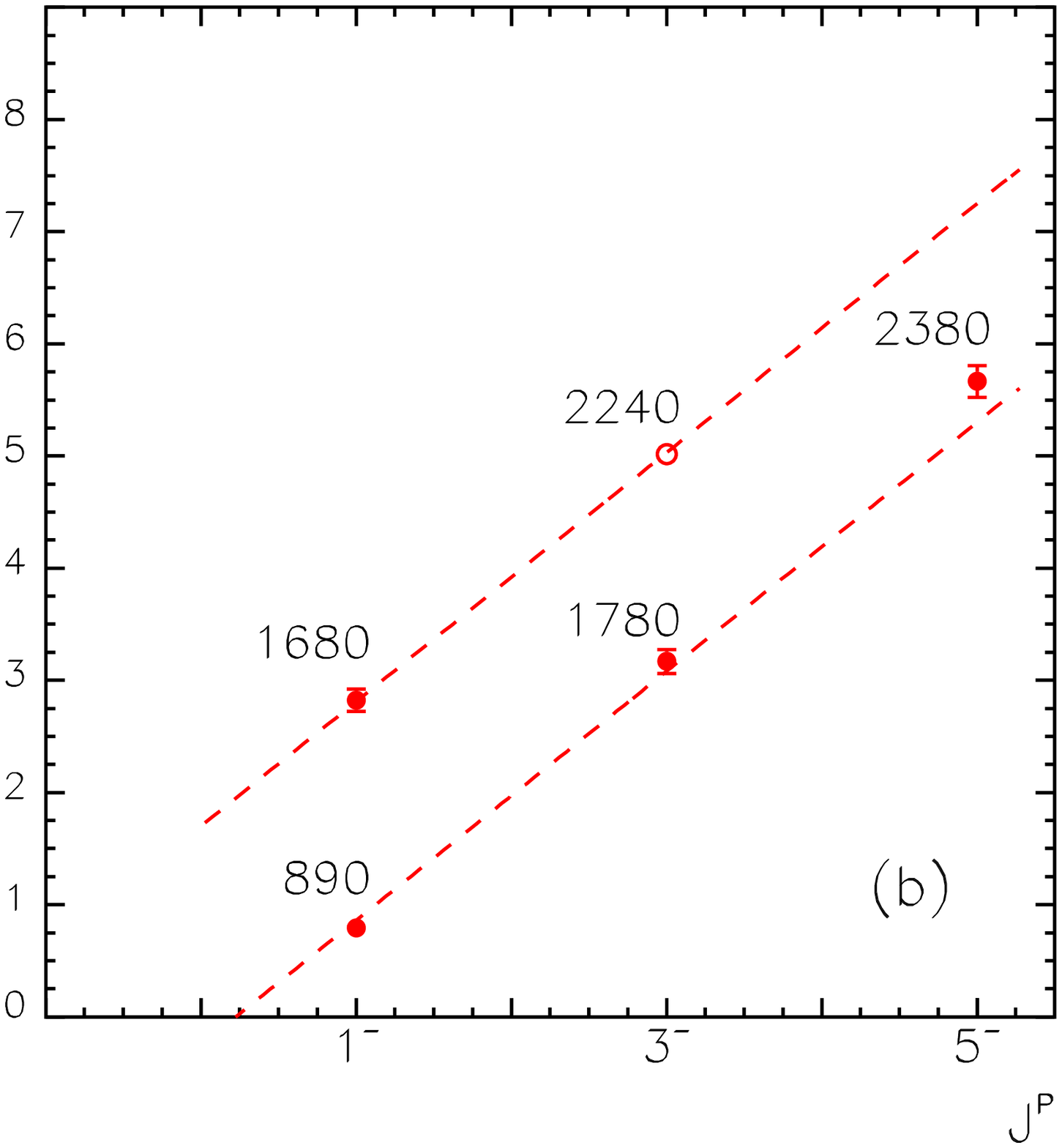,width=7cm}\hspace{-1.5cm}
            \epsfig{file=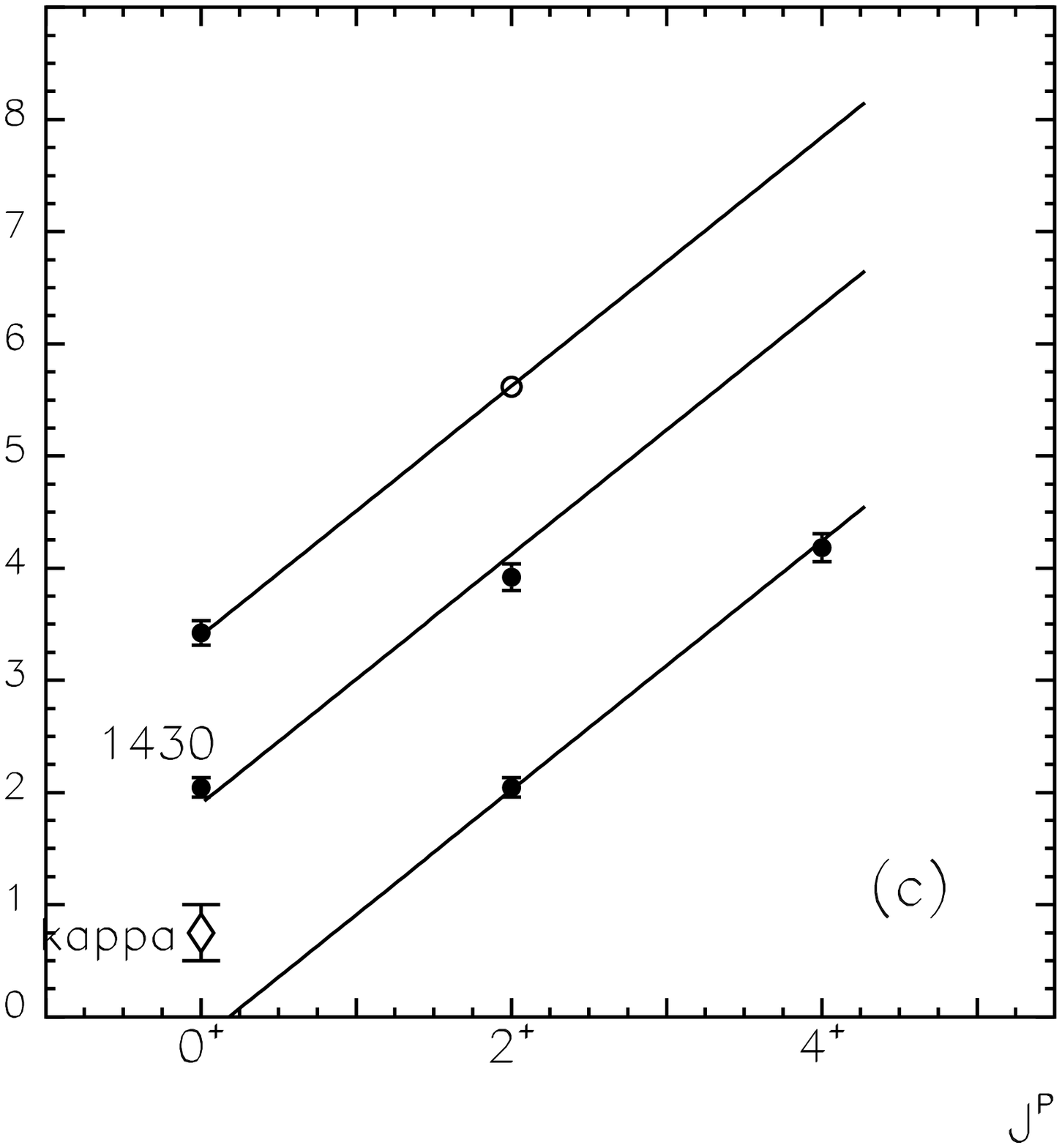,width=7cm}\hspace{-7.13cm}
            \epsfig{file=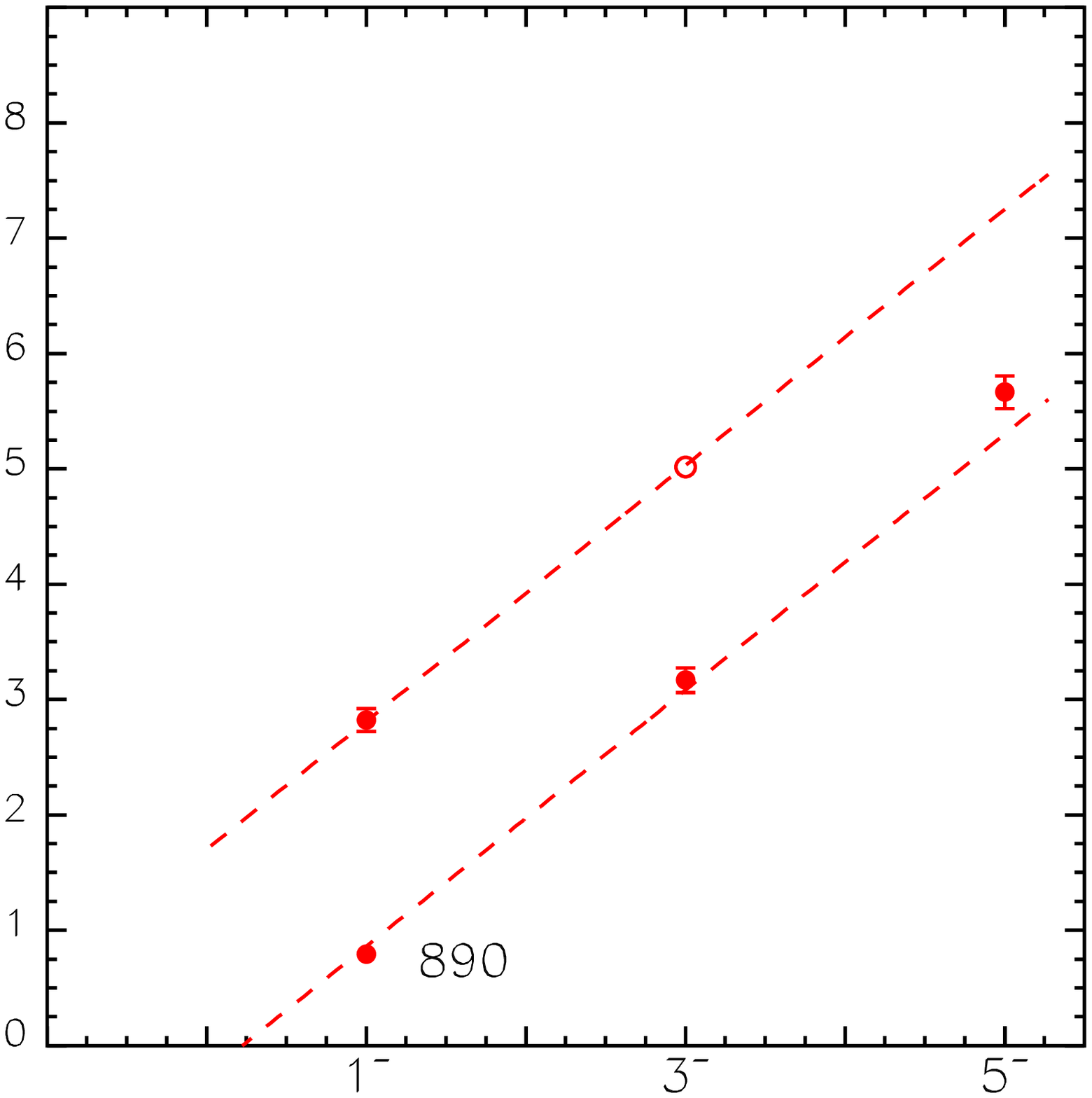,width=7cm}}
\vspace{-0.5cm}
\caption{Trajectories on the $(J,M^2)$ plane, $K$-meson sector.}
\end{figure}

 One may conclude that
there are extra $00^{++}$
states with respect to the $q\bar q$ systematics on the
$(n,M^2)$ and $(J,M^2)$ trajectories:\\
1) the broad state $f_0(1200-1600)$ (it follows from the $K$-matrix
analysis that it is the  descendant of the glueball),\\
2) the light $\sigma$-meson, $f_0(300-500)$ (if it exists),\\
3) $\kappa$-meson, $K_0(700-900)$ (if it exists).

In the pseudoscalar-isoscalar sector $(IJ^{PC}=00^{-+})$,
the situation in the mass region 1400-1800 MeV is rather uncertain, and
one cannot exclude the states existing here, which are superfluous for
the $q\bar q$ trajectories.

$\,$\\

{\bf 2. The $K$-matrix analysis of the ($IJ^{PC}=00^{++}$)-wave.}\\
The trajectory assignment does not specify the content of the state.
Each meson/resonance is a mixture of different components, and their
wave functions are the Fock columns.
For example, $a_0$ and $f_0$ may be considered as:

\centerline{
\begin{tabular}{cc}
$
a_0=\left ( \begin{array}{c}
q\bar q  \\
qq\bar q\bar q  \\
\eta\pi,K\bar K, ...
\end{array}
\right ),
$
&
$
f_0=\left ( \begin{array}{c}
q\bar q  \\
qq\bar q\bar q  \\
\pi\pi,K\bar K, \eta\eta, ...  \\
gg
\end{array}
\right ) \, .
$
\\
\end{tabular}
}

In bare states, determined by the $K$-matrix analysis, the long-range
hadronic components are excluded that makes possible more reliable
determination of the quark-antiquark and gluonium ($gg$) components.

In the $K$-matrix analysis the partial wave amplitude reads:

\centerline{
\begin{tabular}{ccc}
$
\widehat A(s)\ =\ \widehat K(s)
\left[1-i\hat\rho\widehat K(s)\right]^{-1},
$
&
$
\widehat K=\left | \begin{array}{c}
K_{11}, K_{12}, ... \\
K_{21}, K_{22}, ... \\
 ... \\
\end{array}
\right |,
$
&
$
\widehat \rho=\left | \begin{array}{c}
\rho_1, 0, ... \\
0\;\;\;, \rho_2, ... \\
 ... \\
\end{array}
\right |   \, .
$
\\
\end{tabular}
}

For the $00^{++}$ wave, which was analysed in \cite{EPJA},
the following five channels were taken into account:
$1=\pi\pi$, $2=K\bar K$,

$3=\eta\eta$, $4=\eta\eta'$,
$5=\pi\pi\pi\pi$, with the two-particle phase spaces determined as
$$
\rho_{\pi\pi}=\sqrt{\frac{s-4m_\pi^2}{s}}, \quad
\rho_{KK}=\sqrt{\frac{s-4m_K^2}{s}}, \quad ...
$$
while $\rho_{\pi\pi\pi\pi}$ was considerd as either two-$\sigma$
or two-$\rho$ phase space factor.

The fitting parameters are the $K$-matrix
elements, which are represented as the sums of pole terms,
$g^{(n)}_ag^{(n)}_b/(\mu^2_n-s)$, and a smooth $s$-dependent term
$f_{ab}(s)$:
$$
K_{ab}\ =\ \sum \limits_n \frac{g^{(n)}_ag^{(n)}_b}{\mu^2_n-s}
+f_{ab}(s)\ ,
$$
$M_n$ being the masses of bare states and
$g_a$, $g_b$ their couplings.

The combined $K$-matrix analyses of the spectra have been performed for
the mass interval
$ 280\le M \le 1950$ MeV by including
the following final states:
\ba
&I=0:&  \quad \pi\pi,\; \eta\eta,\; K\bar K,\; \eta\eta',\;
\pi\pi\pi\pi\; (\rho\rho,\; \sigma\sigma)\;\cite{EPJA},\nonumber \\
&I=1:&  \quad \pi\eta,\; K\bar K\;\cite{UFN},\nonumber \\
&I=\frac 12:& \quad K\pi; \;\cite{Alex}.\nonumber
\ea
Note that these combined
$K$-matrix analyses have their predecessors published in
\cite{previous}. The necessity of a combined analysis owes to the
existence of large interference effects "resonance--background" as well
as the effects associated with the resonance overlapping. In a
situation of such a type, only a combined fitting to a large number of
reactions allows one to expect reliable results.

Previous analysis \cite{UFN} carried out in 1997--1998
was based on the experimental data as follows:\\
(1) GAMS data on the $S$-wave two-meson  production in the reactions
$\pi p\to \pi^0\pi^0 n$, $\eta\eta n$ and $\eta\eta' n$
at small nucleon momenta transferred, $|t|<0.2$ (GeV/$c$)$^2$
\cite{Alde,Binon};\\
(2) GAMS data on the $\pi\pi$ $S$-wave production in the reaction
$\pi p\to \pi^0\pi^0 n$ at large momentum transfers squared,
$0.30<|t|<1.0$ (GeV/$c$)$^2$ \cite{Alde};\\
(3) BNL data on the reaction $\pi^- p\to K\bar K n$ \cite{Lind};\\
(4) CERN-M\"unich data on $\pi^+\pi^- \to\pi^+\pi^-$ \cite{Grayer};\\
(5) Crystal Barrel data on $p\bar p$ (at rest, from liquid
$H_2$)$\to \pi^0\pi^0\pi^0$, $\pi^0\pi^0\eta$, $\pi^0\eta\eta$
\cite{Anis_Arm,Amsler}.

Now the experimental basis has much broadened, and
additional samples of data are included into the analysis \cite{EPJA}
of the $00^{++}$ wave as follows:\\
(6) Crystal Barrel data on proton-antiproton annihilation in gas:
$p\bar p$ (at rest, from gaseous
$H_2$) $\to \pi^0\pi^0\pi^0$, $\pi^0\pi^0\eta$ \cite{abele}.

One should keep in mind that in liquid hydrogen the $p\bar p$
annihilation is going dominantly from the
$S$-wave state, while in  gas
there is a considerable admixture of the $P$-wave, thus giving us
an opportunity to analyse  the three-meson Dalitz plots in more
detail.\\
(7) Crystal Barrel data on proton-antiproton annihilation in liquid:
$p\bar p$ (at rest, from liquid $H_2$)$ \to
\pi^+\pi^-\pi^0$,
$K^+K^-\pi^0$, $K_SK_S\pi^0$, $K^+K_S\pi^-$ \cite{abele};\\
(8) Crystal Barrel data on neutron-antiproton annihilation in
liquid deuterium: $n\bar p$(at rest, from liquid
$D_2$)$\to \pi^0\pi^0\pi^-$, $\pi^-\pi^-\pi^+$,
$K_SK^-\pi^0$, $K_SK_S\pi^-$ \cite{abele}.

These data allowed us to perform  more confident study of the
two-kaon
channels as compared to what had been done
before.  This is important for the conclusion about the
quark-gluon content of scalar--isoscalar $f_0$-mesons under
investigation.\\
(9) E852 Collaboration data on the $\pi\pi$ $S$-wave production in the
reaction $\pi^-p\to \pi^0\pi^0n$ at the nucleon momentum transfers
squared $0<|t|<1.5 \; ({\rm GeV/c})^2$ \cite{Gunter}.

Experimental data of the E852 Collaboration on the reaction
$\pi^- p \to \pi^0\pi^0n$  at $p_{lab}=18$ GeV/c \cite{Gunter}
together with the GAMS data on the reaction $\pi^- p \to \pi^0\pi^0n$
at $p_{lab}=38$ GeV/c \cite{Alde} give us a solid ground for the
study of  the resonances $f_0(980)$ and $f_0(1300)$, for at large
momenta transferred to the nucleon, $|t| \sim (0.5 - 1.5)$
(GeV/c)$^2$, the production of resonances is accompanied by a small
background, thus allowing us to fix reliably their masses and widths.

The most important ingredients of the new analysis \cite{EPJA} are:\\
(i) the study of $K\bar K$-spectra that led to the determination of the
flavour-octet and flavour-singlet components;\\
(ii) the study of $f_0(1300)$ produced without background:
$\pi N\to \pi\pi N$ at large $|t|$.

Figure 9 taken from \cite{EPJA} demonstrates the
complex-$M$ plane for the $00^{++}$
sector. Here the masses and total widths of resonances are determined by
the position of amplitude poles, $det|1-i\hat\rho\hat K|=0$, the decay
couplings are determined by the pole residues.

The movement of poles in the
complex-$M$ plane
for the states $f_0(980)$, $f_0(1300)$, $f_0(1500)$,
$f_0(1750)$, $f_0(1200-1600)$ with a uniform onset of the decay
channels is shown in Fig. 10.
Technically, to switch on/off the decay channels
for

\begin{figure}
%Fig. 9
\centerline{\epsfig{file=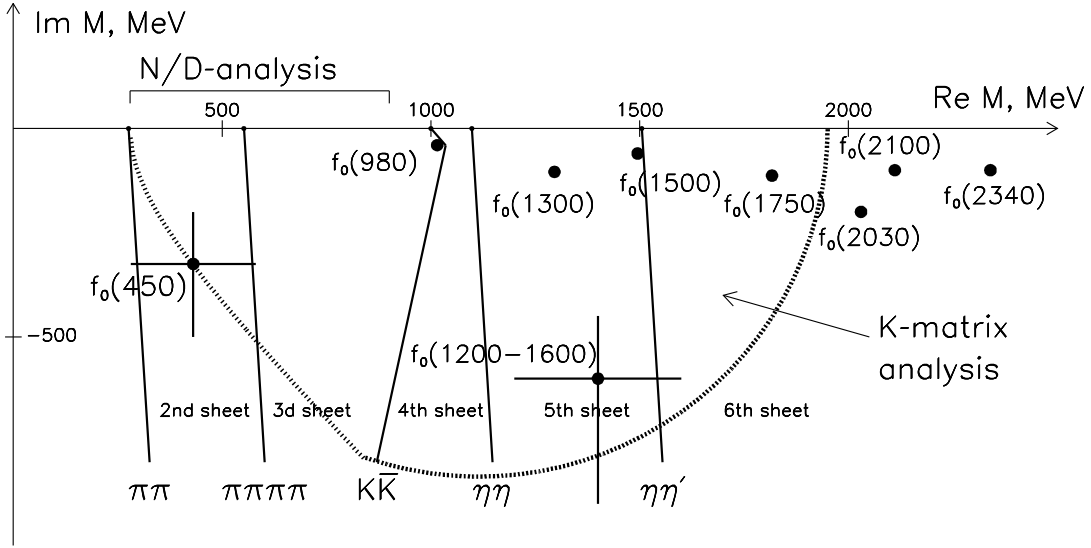,width=15cm}}
\vspace{1cm}
\caption{Complex-$M$ plane in the $(IJ^{PC}=00^{++})$
sector: masses and total widths of resonances are determined by the
position of the amplitude poles $det|1-i\hat\rho\hat K|=0$,
decay couplings are determined by the pole residues.
Dashed line encircles the part of the plane where the $K$-matrix
analysis reconstructs the analytical $K$-matrix amplitude: in this area
the poles corresponding to resonances
$f_0(980)$, $f_0(1300)$, $f_0(1500)$,
$f_0(1750)$ and the broad state $f_0(1200-1600)$ are
located. On the border of this area
 the light $\sigma$-meson denoted as
$f_0(450)$ is shown (however, we have also rather good solution without
$\sigma$-meson pole).}
\end{figure}
\noindent
the  $K$-matrix amplitude
one should substitute in
the $K$-matrix elements
$g^{(n)}_a \to \xi_n(x)g^{(n)}_a$ and
$ f_{ab} \to\xi_f(x)f_{ab}$,
where the parameter-functions
$\xi_n(x)$ and $\xi_f(x)$  satisfy the
following  constraints:
$\xi_n(0)=\xi_f(0)=0$ and $\xi_n(1)=\xi_f(1)=1$, and $x$ varies
in the interval $0\le x\le1$. Then, at $x=0$, the amplitude $\wh A$
turns into the $K$-matrix,
 $\wh A(x\to0)\to\wh K$, and the amplitude poles occur on the
real axis, that corresponds to  stable $f^{bare}_0$-states.

\begin{figure}[h]
%Fig. 10
\centerline{\epsfig{file=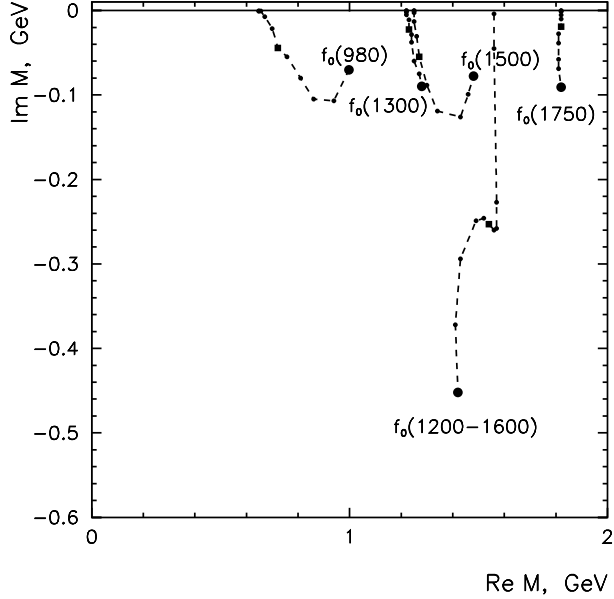,width=9cm}}
\caption{Complex $M$ plane: trajectories of poles
corresponding to the states $f_0(980)$, $f_0(1300)$, $f_0(1500)$,
$f_0(1750)$, $f_0(1200-1600)$ within a uniform onset of the decay
channels.}
\end{figure}

In Fig. 10, one can see  gradual transformation of  bare states into
 real mesons as follows:
\ba
&f^{bare}_0(700\pm100) \to f_0(980), \qquad
&f^{bare}_0(1220\pm40)  \to  f_0(1300), \nonumber \\
&f^{bare}_0(1230\pm40) \to f_0(1500)  \qquad
&f^{bare}_0(1580\pm40)  \to  f_0(1200-1600) \nonumber \\
&f^{bare}_0(1800\pm40) \to f_0(1750) \qquad
&\nonumber
\ea

Although in the $K$-matrix analysis
the pole position of the broad state $ f_0(1200-1600)$ is
defined with large errors, this pole is necessary in all the $K$-matrix
solutions. Also in all solutions the
requirement of the factorization of coupling constants is fulfilled:
$$g_{in}\;\frac 1{M^2-s-i\Gamma M}\;g_{out}.$$

The onset of the decay channels can be illustrated with an example of
the  $f_0$-levels in the potential well (Fig. 11): stable levels
correspond to bare states (Fig. 11a), while overlapping resonances
correspond to real mesons (Fig. 11b).

Figure 12a demonstrates that bare states also form linear
trajectories. Here one can see two $f_0^{bare}$ trajectories as
well as $a_0^{bare}$ and $K_0^{bare}$ ones: all these trajectories have
almost the same slopes as the trajectories of real
resonances, which are also
shown in Fig. 12b for the comparison.

To fix the nonet of bare states by using hadronic decay processes
 two parameters are needed only, namely,
$g,\; \varphi$,
where $g$ is a universal decay coupling and $\varphi$
the mixing angle for $n\bar n$ and $s\bar s$ components in
$f_0^{bare}(1)$ and $f_0^{bare}(2)$.
Decay couplings depend also on the suppression parameter for
the strange quark production probability $\lambda=0.6\pm0.2$, but its value
is fixed by other reactions, for example,
 see \cite{klempt}.

Let us emphasize that these two parameters,
$g$ and $\varphi$, allowed us to describe ten
decay reactions such as
$$
f_0(1) \to \pi\pi,K\bar K,\eta\eta, \quad
f_0(2) \to \pi\pi,K\bar K,\eta\eta,\eta\eta', \quad
K_0 \to \pi K, \quad a_0\to K\bar K,\eta\pi.
$$
The constraints for decay couplings
together with the placement of $f_0^{bare}$ to linear $(n,M^2)$
trajectories
rigidly determine two nonets with $n=1$ and $n=2$.  The extra state,
which was found and investigated
in \cite{EPJA,UFN,previous}, namely, $f_0^{bare}(1580)$, should be
identified as a scalar glueball, for its decay couplings satisfy the
requirements inherent in the gluonium, see discussion in \cite{ufn02}.
So,
$$
f_0^{bare}(1580\pm 40)\to glueball\, .
$$

\begin{figure}
%Fig. 11
\centerline{\epsfig{file=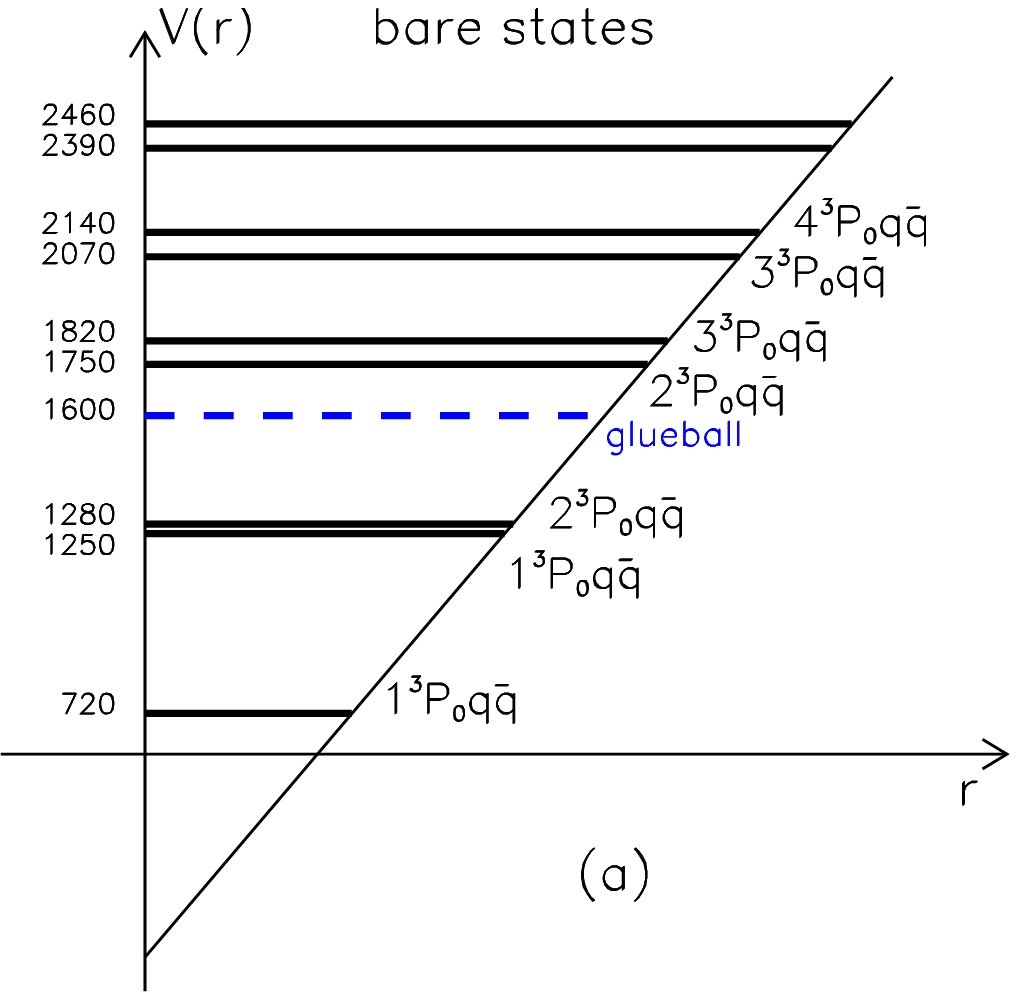,width=8cm}
            \epsfig{file=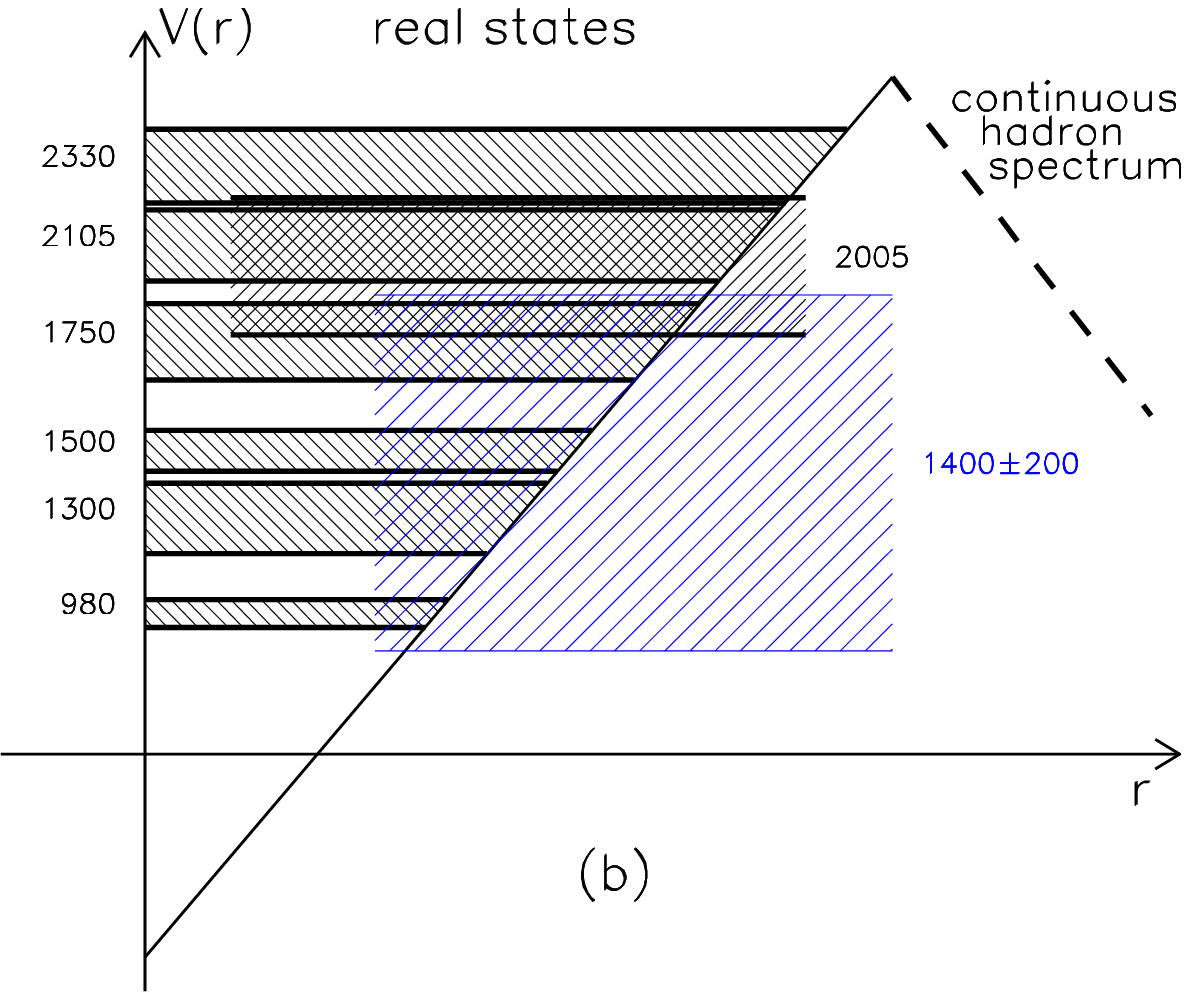,width=8cm}}
\caption{The $f_0$-levels in the potential well depending on
the onset of the decay channels: bare states (a) \\ and real resonances
(b).}
\end{figure}

\begin{figure}[h]
%Fig. 12
\centerline{\epsfig{file=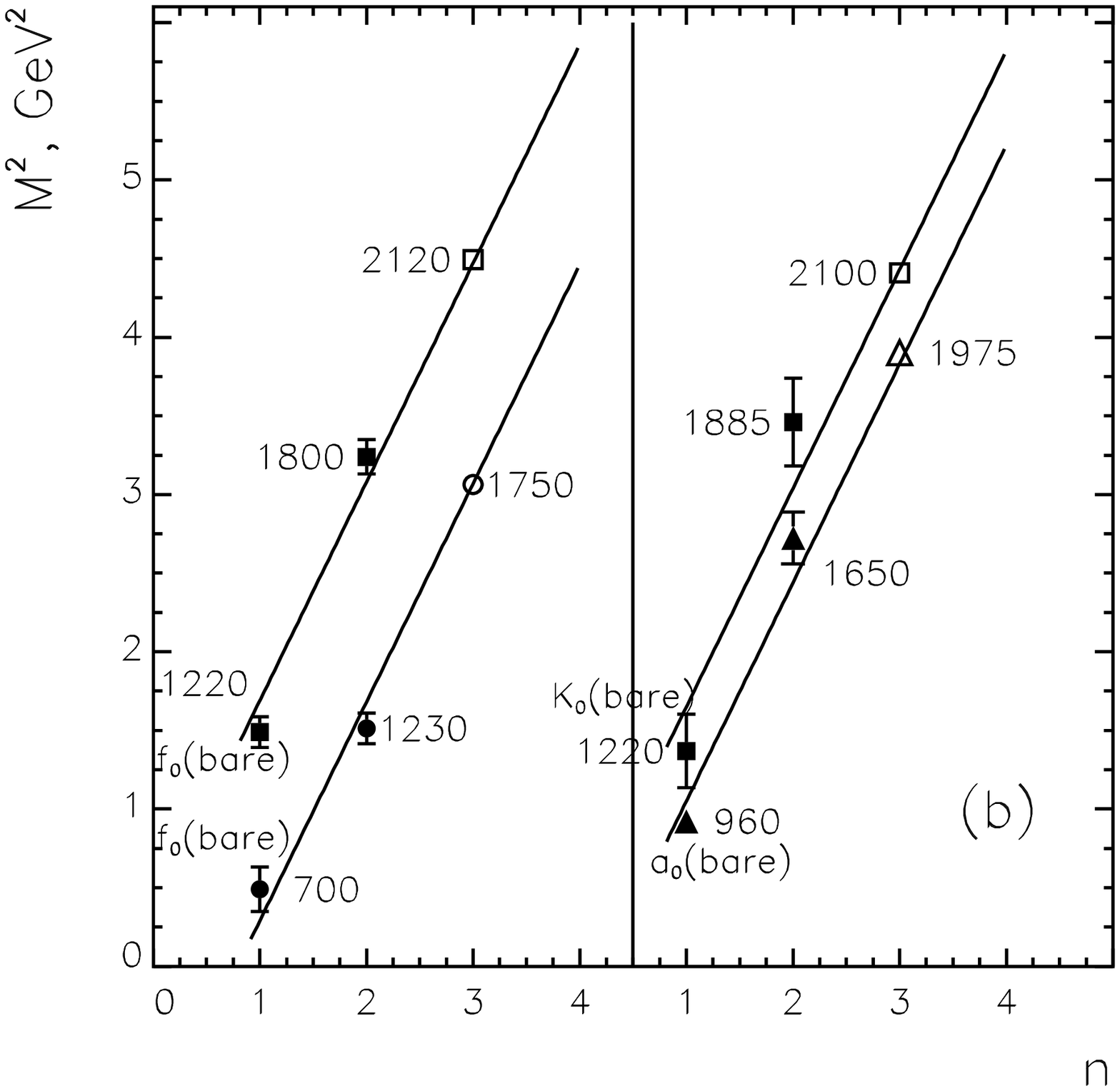,width=7.0cm}
            \epsfig{file=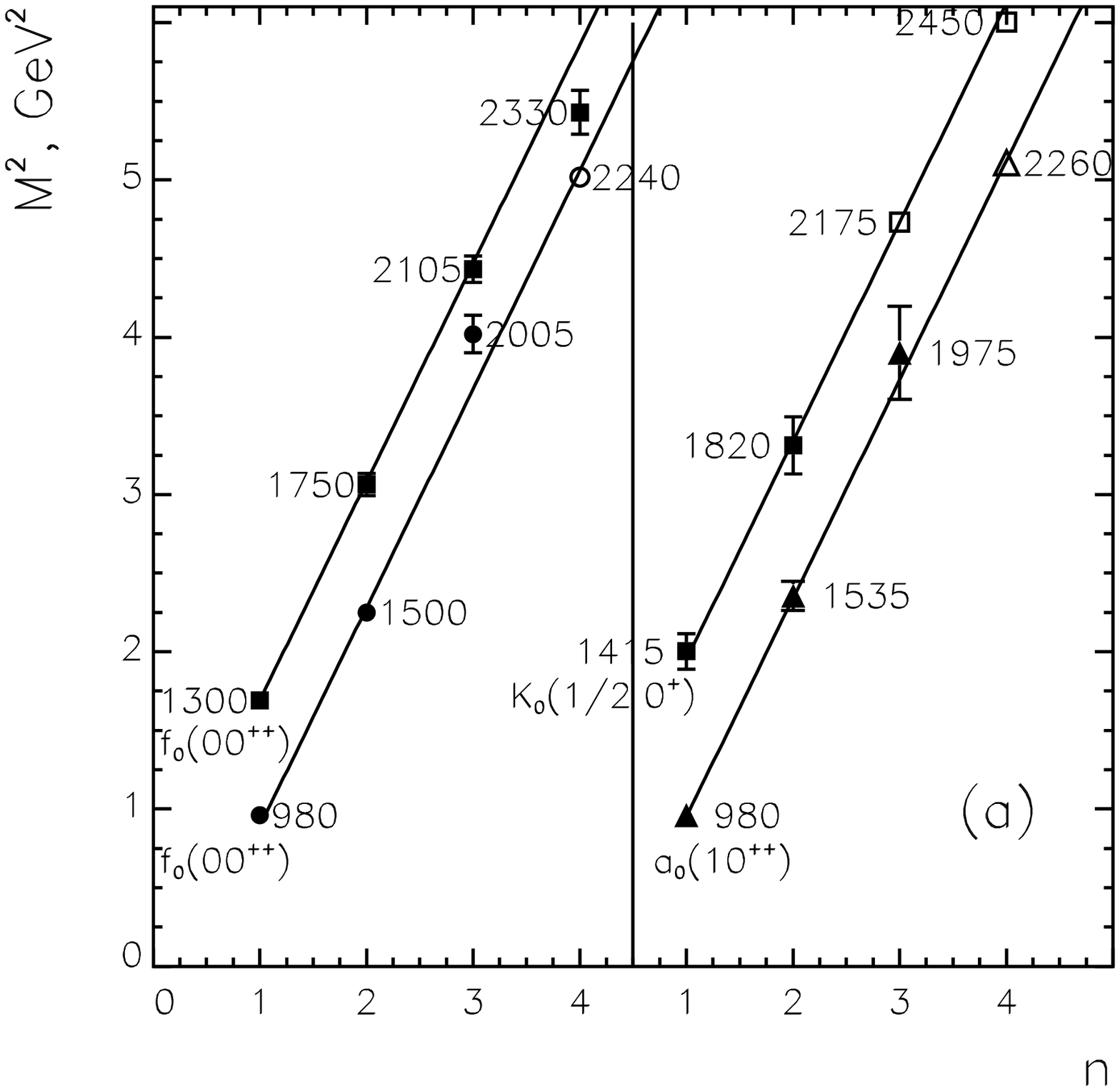,width=7.0cm}}
\vspace{-0.5cm}
\caption{Linear trajectories on the
$(n,M^2)$ plane for bare states (a) and resonances (b).}
\end{figure}
$\,$\\

{\bf 3. Where is the lightest pseudoscalar glueball?}\\
In the pseudoscalar sector, in the mass region under discussion,
one may expect the lightest pseudoscalar glueball, though the
opinions about the mass of pseudoscalar glueball are rather different.
According to lattice gluodynamic calculations, the mass of the lightest
pseudoscalar glueball coincides with that of the tensor glueball, that
is, it must be in the range 2100--2600 MeV \cite{Bali}, while,
according to \cite{Faddeev}, its mass is close to that of the
lightest scalar glueball: 1300--1700 MeV. The plausible existence of
the light $0^{-+}$ glueball looks nice, for it might explain a
considerable production of the $0^{-+}$ states in the radiative
$J/\psi$ decay, in particular, $J/\psi \to \gamma \eta'$ (according to
\cite{Bugg-Mel}, the admixture of the gluonium component in $\eta'$
may be rather large, about 10\%--20\%).  However, among narrow
resonances one cannot see the candidates for pseudoscalar glueball:
$\eta (1295)$ and $\eta (1440)$ lay on linear $q\bar q$ trajectories 
(Fig.  4c), though in \cite{Alexei} it was estimated that the value of 
$gg$ component
in the $\eta(1440)$ can be not small, $20\%\pm 20\%$.  Still, it is
possible that pseudoscalar glueball, after the onset of decay channels,
turned into the broad state in the region 1400--1500 MeV (as it
occurred with scalar one) --- the experimental data do not contradict
this suggestion, see \cite{ufn95} and \cite{Bai}.

I am indebted to A.V. Anisovich, D.V. Bugg, L.G. Dakhno, V.A. Nikonov,
A.V. Sarantsev for the interest to the problem under discussion.
The paper is supported by RFFI grant N 01-02-17861.

\end{document}